\documentclass[preprint,nofootinbib,superscriptaddress]{revtex4-1}

\usepackage{amsmath,amssymb}
\usepackage{graphicx,color,xcolor}
\usepackage{verbatim}
\usepackage{amsfonts}
\usepackage{epstopdf}
\usepackage{comment}
\usepackage{subfig}
\usepackage{physics}
\usepackage{bm}
\usepackage[version=3]{mhchem} 
\usepackage{hyperref}
\usepackage[capitalize]{cleveref}

\newtheorem{thm}{\protect\theoremname}

\newtheorem{prop}[thm]{\protect\propositionname}

\newtheorem{assumption}[thm]{Assumption}

\providecommand{\corollaryname}{Corollary}
\providecommand{\lemmaname}{Lemma}
\providecommand{\propositionname}{Proposition}
\providecommand{\remarkname}{Remark}
\providecommand{\theoremname}{Theorem}

\usepackage[english]{babel}
  \providecommand{\lemmaname}{Lemma}
\providecommand{\theoremname}{Theorem}

\usepackage{algpseudocode}
\usepackage{algorithm}
\usepackage{algorithmicx}

\newcommand{\LL}[1]{\textcolor{red}{[LL: #1]}}

\global\long\def\R{\mathbb{R}}

\global\long\def\ra{\rightarrow}

\global\long\def\Tr{\mathrm{Tr}}

\newcommand{\mc}[1]{\mathcal{#1}}

\newcommand{\wt}[1]{\widetilde{#1}}

\newcommand{\Or}{\mathcal{O}}

\newcommand{\CC}{\mathbb{C}}

\newcommand{\commentout}[1]{}

\begin{document}

\title[L-DMET]{Enhancing  robustness and efficiency of density matrix embedding theory via semidefinite programming and local correlation potential fitting}

\author{Xiaojie Wu}
\affiliation{Department of Mathematics, University of California, 
Berkeley, California 94720, United States}

\author{Michael Lindsey}
\affiliation{Courant Institute of Mathematical Sciences, New York University, New York, NY 10012, United States}

\author{Tiangang Zhou}
\affiliation{School of Physics, Peking University, Beijing, 100871, China}

\author{Yu Tong}
\affiliation{Department of Mathematics, University of California, 
Berkeley, California 94720, United States}

\author{Lin Lin}
\email{linlin@math.berkeley.edu}
\affiliation{Department of Mathematics, University of California, 
Berkeley, California 94720, United States}
\affiliation{Computational Research Division, Lawrence Berkeley 
National Laboratory, Berkeley, California 94720, United States}

%
%
%
%
%

\begin{abstract}
  Density matrix embedding theory (DMET) is a powerful quantum embedding method for solving strongly correlated quantum systems. Theoretically, the performance of a quantum embedding method should be limited by the computational cost of the impurity solver. However, the practical performance of DMET is often hindered by the numerical stability and the computational time of the correlation potential fitting procedure, which is defined on a single-particle level. Of particular difficulty are cases in which the effective single-particle system is gapless or nearly gapless. To alleviate these issues, we develop a semidefinite programming (SDP) based approach that can significantly enhance the robustness of the correlation potential fitting procedure compared to the traditional least squares fitting approach. We also develop a local correlation potential fitting approach, which allows one to identify the correlation potential from each fragment independently in each self-consistent field iteration, avoiding any optimization at the global level. We prove that the self-consistent solutions of DMET using this local correlation potential fitting procedure are equivalent to those of the original DMET with global fitting. We find that our combined approach, called L-DMET, in which we solve local fitting problems via semidefinite programming, can significantly improve both the robustness and the efficiency of DMET calculations. We demonstrate the performance of L-DMET on the 2D Hubbard model and the hydrogen chain. We also demonstrate with theoretical and numerical evidence that the use of a large fragment size can be a fundamental source of numerical instability in the DMET procedure.
\end{abstract}

\date{\today}
\maketitle

\section{Introduction}

In order to treat strong correlation effects beyond the single-particle level for large systems, highly accurate numerical methods such as full configuration interaction (FCI)~\cite{knowles1984new,olsen1990passing,vogiatzis2017pushing}, exact diagonalization (ED)~\cite{lin1990exact,lauchli2011ground}, or the density matrix renormalization group (DMRG)~\cite{White92} with a large bond dimension are often prohibitively expensive. Quantum embedding theories~\cite{sun2016quantum,imada2010electronic,ayral2017dynamical}, such as the dynamical mean field theory (DMFT)~\cite{metzner1989correlated, georges1992numerical,georges1996dynamical,kotliar2006electronic,zhu2019efficient} and density matrix embedding theory (DMET)~\cite{DMET2012, DMET2013, tsuchimochi2015density, bulik2014density, wsj16, cui2019efficient,sun2019finite,cui2020ground},  
offer an alternative approach for treating strongly correlated systems. The idea is to partition the global system into several ``impurities'' to be treated accurately via a high-level theory (such as FCI/ED/DMRG), and to ``glue" the solutions from all impurities via a lower-level theory. This procedure is performed self-consistently until a certain consistency condition is satisfied between the high-level and low-level theories. The self-consistency condition is particularly important when the physical system undergoes a phase transition not predicted by mean-field theory (i.e., the mean-field theory incorrectly predicts the order parameter), and quantum embedding theories provide systematic procedures to qualitatively correct the order parameter.

In this paper we focus on DMET, which has been successfully applied to compute phase diagrams of a number of strongly correlated models, such as the one-band Hubbard model both with and without a superconducting order parameter~\cite{DMET2012, bulik2014density, chen2014intermediate, boxiao2016, Zheng2017, zheng2017stripe}, quantum spin models~\cite{Fan15, gunst2017block}, and prototypical correlated molecular problems~\cite{DMET2013, wsj16, Pham18}. The self-consistency condition is usually defined so that the 1-RDMs obtained from the low-level and high-level theories match each other according to some criterion, such as matching the 1-RDM of the impurity problem~\cite{DMET2012}, matching on the fragment only~\cite{DMET2013,tsuchimochi2015density}, or simply matching the diagonal elements of the density matrix (i.e., the electron density)~\cite{bulik2014density}.  Self-consistency can be achieved  by optimizing a single-body Hamiltonian, termed the correlation potential, in the low-level theory.  Each optimization step requires diagonalizing a matrix, similarly to the self-consistent field (SCF) iteration step in the solution of the Hartree-Fock equations.
 
However, the correlation potential optimization step can become a computational bottleneck, even compared to the cost of of the impurity solvers. This is because in DMET, the size of each impurity is often thought of as a constant, and therefore the cost for solving all of the impurity problems always scales linearly with respect to the global system size. Meanwhile, the correlation potential fitting requires repeated solution of problems at the single-particle level and is closely related to the density inversion problem~\cite{JensenWasserman2017,WuYang2003}.  In order to evaluate the derivative, the computational effort is similar to that of a density functional perturbation theory (DFPT) calculation \cite{BaroniGironcoliDalEtAl2001}. The number of iterations to optimize the correlation potential can also increase with respect to the system size, especially for gapless systems, provided the procedure can converge at all. 

In this paper, we propose two improvements to significantly increase the efficiency and the robustness of the correlation potential fitting procedure. To enhance the robustness, we propose to reformulate the correlation potential fitting problem as a semidefinite program (SDP). It is theoretically guaranteed that when the correlation potential is uniquely defined, it coincides with the optimal solution of the SDP. Moreover, as a convex optimization problem, the SDP has no spurious local minima. To improve the efficiency, we introduce a local correlation potential fitting approach. The basic idea is to perform local correlation potential fitting on each impurity to match the high-level density matrix and the local density matrix. Then the local correlation potentials are patched together to yield the high-level density matrix. We may further combine the two approaches and utilize the SDP reformulation for each impurity. This approach is dubbed local-fitting based DMET (L-DMET). We prove that the results obtained from DMET and L-DMET are equivalent. Nonetheless, L-DMET scales linearly with respect to the system size in each iteration of DMET. 
It is numerically observed that L-DMET does not require more iterations than DMET. 
This is particularly advantageous for the simulation of large systems. 



The rest of the paper is organized as follows. In Section \ref{sec:dmet_description}, we first briefly present the formulation of DMET. In particular, DMET can be concisely viewed from a linear algebraic perspective using the CS decomposition. The SDP reformulation of the correlation potential fitting is introduced in Section \ref{sec:convex} as an alternative approach to the least squares problem in DMET. In Section \ref{sec:l-dmet}, we present the local correlation fitting approach (L-DMET) and show the equivalence between the fixed points of DMET and L-DMET. The relation between the current work and a few related works, such as the finite temperature generalization and the p-DMET \cite{WuCuiTongEtAl2019}, is discussed in Section \ref{sec:otherconsider}. Numerical results for the 2D Hubbard model and the hydrogen chain are given in Sections \ref{sec:2d_hubbard} and  \ref{sec:hydrogen}, respectively. We conclude in Section \ref{sec:conclusion}. The proofs of the propositions in the paper are given in the appendices.

\section{Brief review of DMET}
\label{sec:dmet_description}
Consider the problem of finding the ground state of the quantum many-body Hamiltonian operator in the second-quantized formulation 
\begin{equation}
  \label{eq:13}
  \hat H = \hat t + \hat v^\mathrm{ee} = \sum_{pq}^{L} t_{pq}\hat a^\dagger_p\hat a_q+\frac{1}{2}\sum_{pqrs}^{L}(pr|qs)\hat a^\dagger_p\hat a_q^\dagger\hat a_s\hat a_r.
\end{equation}
Here $L$ is the number of spin orbitals. The corresponding Fock space is denoted by $\mc{F}$, which is of dimension $2^{L}$.   The number of electrons is denoted by $N_e$. We partition the  $L$ sites into $N_\text{f}$ fragments. Without loss of generality, we assume each fragment has the same size $L_A$, though a non-uniform partition  is possible as well. We define the set of block-diagonal matrices with the sparsity pattern corresponding to the fragment partitioning as
\begin{equation}
  \label{eq:3}
  \mc{S} = \left\{A=\bigoplus_{x=1}^{N_{\mathrm{f}}}A_x  \  \Bigg| \,  \,A_x \in \mathbb{C}^{L_A \times L_A}, \, A_x=A_x^\dagger \text{ for } x = 1,\dots,N_{\mathrm{f}}\right\},
\end{equation}
where $\bigoplus$ indicates the direct sum of matrices, i.e.
\begin{displaymath}
\bigoplus_{x=1}^{N_{\mathrm{f}}}A_x=\left(\begin{array}{cccc}
A_{1} & 0 & \cdots & 0\\
0 & A_{2} & \cdots & 0\\
\vdots & \vdots & \ddots & \vdots\\
0 & 0 & \cdots & A_{N_{\mathrm{f}}}
\end{array}\right).
\end{displaymath}

Density matrix embedding theory (DMET) can be formulated in a self-consistent manner with respect to a \textit{correlation potential} $u\in \mc{S}$. For a given $u$,  the low-level (also called the single-particle level) Hamiltonian takes the form
\begin{equation}
  \label{eqn:H_ll}
  \hat H^\text{ll}(u) = \hat f + \hat c(u).
\end{equation}
Here $\hat c(u) = \sum_{pq}u_{pq}\hat a_p^\dagger\hat a_q$ is a quadratic interaction associated with the correlation potential.
When the ground state  of $\hat{H}^{\text{ll}}$ can be uniquely defined, this ground state is a single-particle Slater determinant denoted by $\ket{\Psi^\text{ll}(u)}$, given by a matrix $C\in \CC^{L\times N_e}$.  The associated low-level density matrix is denoted by $D^{\text{ll}}(u):=CC^{\dag}$.
Here $\hat{f}:=\sum_{pq}f_{pq} \hat{a}^{\dag}_p \hat{a}_q$ is given by a fixed matrix $f$. The simplest choice is $f=t$, but other choices are possible as well \cite{wsj16}. Then the low-level density matrix can be expressed as $D^{\text{ll}}(u)=\mc{D}(f+u,N_e)$, which is well-defined when the matrix $f+u$ has a positive gap between the $(N_e)$-th and $(N_e+1)$-th eigenvalues. (Note that throughout we shall use the general notation $\mathcal{D}(h,N)$ to denote the $N$-particle density matrix induced by the non-interacting Hamiltonian specified by the single particle matrix $h$.) 

For each fragment $x$, the Schmidt decomposition of the Slater determinant   $\ket{\Psi^\text{ll}(u)}$  can be used 
to identify a certain subspace $\mathcal{F}_x \subset \mathcal{F}$ that contains  $\ket{\Psi^\text{ll}(u)}$ as follows. Without loss of generality, we assume the fragment $x$ consists of first $L_A$ orbitals
labeled by $\{1,2,\ldots,L_{A}\}$. Since $C$ has orthonormal columns as $C^{\dag}C=I_{N_e}$,  we may apply the CS decomposition \cite{VanLoan1985,GolubVan2013} and obtain
\begin{equation}
C=\left(\begin{array}{c}
U_{A}\Sigma_{A}V^{\dagger}\\
U_{B}\Sigma_{B}V^{\dagger}+U_{\text{core}}V_{\perp}^{\dagger}
\end{array}\right).
\label{eqn:c_ll}
\end{equation}
Here $U_{A}\in \CC^{L_A\times L_A}$, $U_{B}\in \CC^{(L-L_A)\times L_A}$, $U_{\text{core}}\in \CC^{(L-L_A)\times (N_e-L_A)}$,
$V\in\CC^{N_e\times L_A}$ and $V_{\perp}\in \CC^{N_e\times (N_e-L_A)}$ are all column orthogonal matrices. $\Sigma_{A},\Sigma_{B}\in\CC^{L_A\times L_A}$ are non-negative, diagonal matrices and they satisfy $\Sigma_A^2+\Sigma_B^2=I_{N_e}$. Furthermore, $U_{B}^{\dagger}U_{\text{core}}=0$, $V^{\dagger}V_{\perp}=0$.
The CS decomposition \eqref{eqn:c_ll} defines a low-level density matrix. On the other hand, the decomposition as well as $U_A,U_B,U_{\text{core}}$ can be deduced from $D^{\text{ll}}$ directly. The relation is given in Appendix \ref{sec:bath}. 

Throughout the paper, we assume the following condition is satisfied.
\begin{assumption}\label{assump:nondegen}
We assume $N_e>L_A$, and for each fragment $x$, the diagonal entries of $\Sigma_A,\Sigma_B$  in Eq. \eqref{eqn:c_ll} are not $0$ or $1$. 
\end{assumption}
When Assumption \ref{assump:nondegen} is violated, particularly when $L_A$ is large relative to $N_e$ (such as in the context of a large basis set), the choice of the correlation potential is generally not unique (Appendix \ref{app:unique}).

The decomposition \eqref{eqn:c_ll} allows us to define the fragment, bath and core orbitals  
as the columns of
\[
\Phi_{x}^{\mathrm{frag}}=\left(\begin{array}{c}
I_{L_A}\\
0
\end{array}\right), \quad 
\Phi_{x}^{\mathrm{bath}}=\left(\begin{array}{c}
0\\
U_{B}
\end{array}\right),\quad\Phi_{x}^{\text{core}}=\left(\begin{array}{c}
0\\
U_{\text{core}}
\end{array}\right).
\]
In particular, the number of bath orbitals is only $L_A$. This is a key observation in DMET \cite{DMET2012,DMET2013}. The rest of the single-particle orbitals orthogonal to $\Phi_{x}^{\mathrm{frag}},\Phi_{x}^{\mathrm{bath}},\Phi_{x}^{\mathrm{core}}$ are called the virtual orbitals and are denoted by
\[
\Phi_{x}^{\text{vir}}=\left(\begin{array}{c}
0\\
U_{\text{vir}}
\end{array}\right).
\]
The virtual orbitals are not explicitly used in DMET. We also define the set of impurity orbitals, which consists of fragment and bath orbitals, as $$\Phi_{x}=
\begin{pmatrix}
  \Phi_{x}^\mathrm{frag} & \Phi_{x}^\mathrm{bath}
\end{pmatrix}
=\left(\begin{array}{cc}
I_{L_A} & 0\\
0 & U_{B}
\end{array}\right).$$

Using a canonical transformation, the fragment, bath, core and virtual orbitals together allow us to define a new set of creation and annihilation operators $\{\hat{c}_p^\dagger, \hat c_p\}$ in the Fock space  satisfying several properties. 
 First,  $\hat c_1^\dagger, \ldots, \hat c_{L_A}^\dagger$ correspond exactly to  $\hat a_p^\dagger$ for all $p$ in the fragment $x$. Second, the operators $\hat c_1^\dagger, \ldots, \hat c_{2 L_A}^\dagger$ generate an active Fock space $\mathcal{F}^{\mathrm{act}}_x$ of dimension $2^{2L_{A}}$, such that the low-level wavefunction can be written as $\ket{\Psi^\text{ll}(u)} = \ket{\Psi_x^\text{act}(u)} \otimes \ket{\Psi_x^\text{inact}(u)}$, where $\ket{\Psi_x^\text{inact}(u)}$ lies in the inactive space generated by $c_{2 L_A+1}^\dagger, \ldots, c_{N_e}^\dagger$ corresponding to the core orbitals (the virtual orbitals do not contribute to the Slater determinant $\ket{\Psi^\text{ll}(u)}$).  Then the subspace $\mathcal{F}_x$, called the $x$-th impurity space, can be defined by 
\[
  \mathcal{F}_x = \{ \ket{\Psi} \otimes \ket{\Psi_x^\text{inact}(u)} \, : \, \ket{\Psi} \in \mathcal{F}^{\mathrm{act}}_x\}.
\]
Evidently $\ket{\Psi^\text{ll}(u)} \in \mathcal{F}_x \simeq \mathcal{F}^{\mathrm{act}}_x$. Then by a Galerkin projection onto 
$\mathcal{F}_x$ \cite{wsj16}, one derives a ground-state quantum many-body problem on each of the active spaces 
$\mathcal{F}^{\mathrm{act}}_x$, specified by an impurity Hamiltonian (or embedding Hamiltonian) of the following form:
%
\begin{equation}
\label{eq:Hemb}
\hat H^\text{emb}_x= \hat t_x + \hat{v}^\mathrm{emb}_x + \hat v^\mathrm{ee,emb}_x- \mu \hat N_x^\mathrm{frag}.
\end{equation}
Here $\hat{t}_x$ is a single-particle operator specified by the active-space block of the canonically transformed single-particle matrix $t$, $\hat v^\mathrm{ee,emb}_x$ is a two-particle interaction specified by the          active-space block of the canonically transformed two-particle tensor $(pr|qs)$, and $\hat{v}^\mathrm{emb}_x$ is an additional single-particle operator due to the core electron wavefunction $\ket{\Psi_x^\text{inact}(u)}$ in the inactive space. Finally,  $\hat{N}_x^\mathrm{frag}$ is the total number operator for the fragment part of the $x$-th impurity, and $\mu$ is a scalar determined by a criterion to be discussed below.

Given Assumption \ref{assump:nondegen}, the number of core orbital electrons in  $\ket{\Psi_x^\text{inact}(u)}$ is $N_e - L_A$, so the number of electrons in the active space of each impurity is  equal to $L_A$.  
Let  $D_x^\text{hl}\in \CC^{2L_A\times 2L_A}$ be the single-particle density matrix corresponding to the $L_A$-particle ground state of the many-body Hamiltonian $H^\text{emb}_x$, so  $\Tr[D_x^\text{hl}]=L_{A}$. Define the matrix $E = (I_{L_A} \quad 0_{L_A\times L_A})^\top$, 
so the upper-left block of the density matrix $D_x^\text{hl}$, corresponding to the fragment,
can be written as $D_x^\text{hl,frag}:=E^{\top} D_x^\text{hl} E$. Going through all fragments, we obtain the diagonal matrix \textbf{}blocks of the high-level density matrix as
\begin{equation}
D^\text{hl,frag}:=\bigoplus_{x=1}^{N_{\mathrm{f}}}D_x^\text{hl,frag}\in \mc{S}.
\label{eqn:dm_hl_frag} 
\end{equation}
However, the total number of electrons from all fragments must still be equal to $N_e$. This requires the following condition to be satisfied
\begin{equation}
\Tr[D^\text{hl,frag}]=\sum_{x=1}^{N_\mathrm{f}}\Tr(D_x^\text{hl,frag}) = N_e.
\label{eqn:Ne_condition}
\end{equation}
Eq. \eqref{eqn:Ne_condition} is achieved via the appropriate choice of the Lagrange multiplier (i.e., chemical potential)  $\mu$ in the definition~\eqref{eq:Hemb} of the embedding Hamiltonian. 



Once the matrix blocks in $D^\text{hl,frag}$ are obtained, DMET adjusts the correlation potential by solving the following least squares problem 
\begin{equation}
  \label{eqn:least_squares}
  \min_{u \in \mc{S}^{0}} \sum_{x=1}^{N_{\mathrm{f}}} \|D_x^\text{hl,frag}-(\Phi_{x}^\mathrm{frag})^{\dag}\mathcal D(f+u, N_e)\Phi_{x}^\mathrm{frag}\|_F^2.
\end{equation}
Here $(\Phi_{x}^\mathrm{frag})^{\dag}\mathcal D(f+u, N_e)\Phi_{x}^\mathrm{frag}$ gives the the diagonal matrix block corresponding to the $x$-th fragment. We define $\mc{S}^0:=\{A\in\mc{S}~|~\Tr[A]=0\}$, and the traceless condition is added due to the fact that adding a constant in the diagonal entries of $u$ does not change the objective function. The minimization problem \eqref{eqn:least_squares} can be solved with standard nonlinear optimization solvers such as the conjugate gradient method or the quasi-Newton method, and the gradient of the objective function with respect to $u$ can be analytically calculated \cite{wsj16}.  

Finally, in order to formulate the DMET self-consistent loop, we  define the nonlinear mapping $\mathfrak{D}:u\mapsto D^{\text{hl,frag}}$. This mapping takes the correlation potential $u$ as the input, generates the bath orbitals, and solves all impurity problems to obtain the matrix blocks $D^{\text{hl,frag}}$. We also define the mapping $\mathfrak{F}:D^{\text{hl,frag}}\mapsto u$, which takes the high-level density matrix blocks $D^{\text{hl,frag}}$ as the input and updates the correlation potential. Formally, the self-consistency condition of DMET can be formulated as
\begin{equation}
  \label{eqn:dmet_selfconsistency}
  u=\mathfrak{F}\circ \mathfrak{D}(u).
\end{equation}

In the discussion above, the definition of the mapping $\mathfrak{F}$ and the well-posedness of the nonlinear fixed point problem hinges on the uniqueness of the solution of Eq. \eqref{eqn:least_squares}. In Appendix \ref{app:unique} we show that the condition $N_e \geq L_A$ as in Assumption \ref{assump:nondegen} is a necessary condition for the correlation potential to be uniquely defined. The practical consequences of this assumption will also be studied in Section \ref{sec:hydrogen}.

\section{Enhancing the robustness: Semidefinite programming}
\label{sec:convex}
In order to improve the robustness of correlation potential fitting, we develop an alternative approach to the least squares approach in Eq. \eqref{eqn:least_squares}. 
Consider a mapping $F:\mathcal{S}\ra\R$ defined by 
\[
F(u)=\mc{E}_{N_{e}}[f+u],
\]
 where  $\mc{E}_{N_{e}}$ gives  the sum of the lowest $N_{e}$ eigenvalues of the matrix $f+u$. Note that $\mc{E}_{N_e}$ is a concave function, and   $F$ is
a composition of a concave function with a linear function. Hence 
$F$ is a concave function on $\mc{S}$. However, $F$ is not smooth: there
are singular points where $f+u$ is gapless, i.e., there is no gap between the 
$(N_e)$-th and $(N_e+1)$-th eigenvalues.

Whenever a matrix $A$ is gapped, we have $\nabla_A \mc{E}_{N_e}(A) = \mathcal{D}(A,N_e)$.
This is in fact a slight generalization of the Hellmann-Feynman theorem, which is precisely the case when $N_e = 1$. Therefore $\nabla_{u_x}F(f+u) = (\Phi_{x}^\mathrm{frag})^{\dag}\mathcal D(f+u, N_e)\Phi_{x}^\mathrm{frag}$ whenever $f+u$ is gapped.

The correlation potential fitting problem  requires us to evaluate the inverse of the gradient mapping $\nabla_u F =  \bigoplus_{x=1}^{N_\mathrm{f}} \nabla_{u_x} F$ at the point $D^{\mathrm{hl},\mathrm{frag}}$. Since $F$ is concave, the inverse mapping relates to the gradient of the concave conjugate, or the Legendre-Fenchel transform~\cite{rock}. 
The conjugate is denoted by $F^*:\mc{S}\to\R$ and defined as
\begin{equation}
  \label{eq:conjugate}
  F^{*}(P)=\inf_{u\in \mc{S}^0}\left\{ \sum_{x=1}^{N_{\mathrm{f}}}\Tr[P_{x}u_{x}]-F(u)\right\}, \quad P\in\mc{S}. 
\end{equation}
Here we use the new notation $P$ to denote a generic block diagonal matrix that may not be the same as $D^{\mathrm{hl},\mathrm{frag}}$.
Again we may restrict $u$ to be within the set $\mc{S}^0$
since the objective function of Eq.~\eqref{eq:conjugate} is invariant under the transformation $u \leftarrow u + \mu I$. 
In fact, the minimization problem in Eq.~\eqref{eq:conjugate} is a slightly generalized formulation of the variational approach for finding the optimal effective potential (OEP)\cite{WuYang2003,JensenWasserman2017}, as well as the Lieb approach for finding the exchange-correlation functional \cite{Lieb1983}.
We will show:
\begin{prop}
\label{prop:convex}
Suppose $0\prec D_{x}^{\mathrm{hl},\mathrm{frag}}\prec I_{L_{A}}$
for $x=1,\ldots,N_{\mathrm{f}}$ and $\sum_{x=1}^{N_{\mathrm{f}}}\Tr[D_{x}^{\mathrm{hl},\mathrm{frag}}]=N_{e}$.
Then the convex optimization problem for the evaluation of $F^*(D^{\mathrm{hl},\mathrm{frag}})$, i.e., the optimization problem in Eq.~\eqref{eq:conjugate} where $P = D^{\mathrm{hl},\mathrm{frag}}$, admits an optimizer $u^{\star}$.  Then $D^{\mathrm{hl},\mathrm{frag}}$
lies in the supergradient set of $F$ at $u^\star$.  If $f+u^{\star}$ has a gap between
its $(N_{e})$-th and $(N_{e}+1)$-th eigenvalues (ordered increasingly),
then $\mathcal{D}(f+u^\star, N_e)$ has diagonal blocks matching $D^{\mathrm{hl},\mathrm{frag}}$,
i.e., we achieve exact fitting. If $f+u^{\star}$ has no gap, then
the ground state and the mapping $\mathcal{D}(f+u^\star, N_e)$ are ill-defined, and, assuming that the optimizer $u^\star$ is unique, there is no correlation
potential $u$ that yields a well-defined low-level density matrix achieving exact fitting.
\end{prop}

The proof of Proposition~\ref{prop:convex} is provided in Appendix~\ref{app:convex}.
We remark that the matter of whether there exists a \emph{unique} optimizer $u^\star \in \mathcal{S}^0$ appears to be subtle. Such uniqueness would follow from the 
strict concavity of $\mathcal{F}\vert_{\mathcal{S}^0}$, if it could be established. 

Now we further demonstrate that the convex optimization problem of Proposition~\ref{prop:convex} can be equivalently
reformulated as a semidefinite program (SDP), which can be tackled
numerically by standard and robust solvers.
The equivalence is established by the following proposition, and the proof is in Appendix~\ref{app:sdp}.
\begin{prop}
\label{prop:sdp}
Optimizers $u^\star$ as in Proposition~\ref{prop:convex} can be obtained from optimizers $(u^\star, Z^\star, \alpha^\star)$
of the semidefinite program 
\begin{equation}
\begin{aligned}
 & \underset{u\in \mc{S}^{0},\,Z\in\mathbb{C}^{L\times L},\,\alpha\in\R}{\mathrm{minimize}}  & \sum_{x=1}^{N_{\mathrm{f}}}\Tr[D_{x}^{\mathrm{hl},\mathrm{frag}}u_{x}]-\alpha N_{e}+\Tr(Z)\\
 &\mathrm{subject\ to} &f+u+Z-\alpha I\succeq 0\\
 &  & Z\succeq 0.
\end{aligned}
\label{eqn:min_sdp}
\end{equation}
\end{prop}

The minimization problem \eqref{eqn:min_sdp} appears to be significantly different from standard problems in electronic structure calculation. However, we may verify that if $u^{\star}$ is a minimizer and $f+u^{\star}$ is gapped with the standard eigenvalue decomposition
\[
(f+u^{\star}) \psi_k=\lambda_k\psi_k,
\]
then $\alpha$ is a chemical potential satisfying $\lambda_{N_e}<\alpha<\lambda_{N_e+1}$, and $Z=\sum_{i=1}^{N_e} (\alpha-\lambda_i)\psi_i\psi_i^{\dag}$. Then $Z\succeq 0$, $f+u^{\star}+Z-\alpha I=\sum_{a=N_e+1}^{L} (\lambda_a-\alpha)\psi_i\psi_i^{\dag}\succeq 0$, and the objective function of Eq. \eqref{eqn:min_sdp} is indeed equal to 
$\sum_{x=1}^{N_{\mathrm{f}}}\Tr[D_{x}^{\mathrm{hl},\mathrm{frag}}u_{x}]-\sum_{i=1}^{N_e}\lambda_i=F^{*}(D^{\mathrm{hl},\mathrm{frag}})$.

Hence, our new approach 
improves upon that of  \eqref{eqn:least_squares} in two ways. First, whenever exact fitting is possible, we can solve the problem with 
more robust optimization algorithms with strong guarantees of success and which are not, in particular, susceptible 
to spurious local minima. Second, whenever exact fitting is impossible, we can certify that this is indeed the case by observing that the correlation potential that 
we obtain defines a gapless system. 
By contrast, if 
exact fitting is not achieved in the least squares approach, it may not be possible to certify that the optimization algorithm is not merely stuck in a local minimum of the objective function.

\section{Enhancing the efficiency: local correlation potential fitting}
\label{sec:l-dmet}

The convex optimization formulation improves the robustness of the correlation potential fitting procedure. However, we still need to solve an  SDP with $(L_A+1)L/2$ variables (the constant $1/2$ is due to the symmetry of the correlation potential), while intermediate variables such as $Z$ can be of size $L\times L$.  Hence for large inhomogeneous systems, the cost of the correlation potential fitting can be significant and may still outweigh the cost of the impurity solver. In this section, we develop a local fitting method, which decouples the global SDP problem into $N_{\mathrm{f}}$ local fitting problems, each of size $L_A\times L_A$ only. The cost of the correlation potential fitting procedure then scales linearly with respect to $L$, assuming the total number of iterations does not increase significantly. 

The idea of performing a local fitting is motivated from the following consideration.
The embedding Hamiltonian $\hat H^\text{emb}_x$ is obtained by a Galerkin projection of $\hat H$ to $\mc{F}_x$ via a canonical transformation of the creation and annihilation operators. We may apply the same transformation to the low-level Hamiltonian $\hat H^{\text{ll}}$, modified by a potential $v_x$ on the fragment, to obtain a quadratic Hamiltonian \begin{equation}
\hat H^{\text{ll,emb}}_x=\sum_{p,q=1}^{2L_A} (\wt{f}_x+E v _xE^{\top})_{pq} c^{\dag}_{p}c_q, \quad v\in\mc{S}.
\label{eqn:h_ll_emb}
\end{equation}
Here $\widetilde f_x = \Phi_x^\dagger(f+u)\Phi_x$ is the projected Fock matrix onto the impurity $x$. As before $E = (I_{L_A} \quad 0_{L_A\times L_A})^\top$, and then $E v_xE^{\top}\in \CC^{2L_A\times 2L_A}$ is defined on the impurity.  When $v_x =0$,  the fragment density matrix obtained from the ground state of $\hat H^{\text{ll,emb}}_x$ should agree with the global low-level density matrix restricted to the same fragment. (This statement will be justified in Appendix \ref{app:equivalency}.) Then instead of the global least squares fitting problem, we may solve a modified least squares problem 
\begin{equation}
  \label{eqn:lsq_local}
  \min_{u\in \mc{S}^0}\sum_{x=1}^{N_{\mathrm{f}}}\|D_x^\text{hl, frag}-E^\top \mathcal D(\widetilde f_x+E v_x E^\top, L_A)E\|_F^2.
\end{equation}
In contrast to Eq.~\eqref{eqn:least_squares}, the minimizations with respect to different matrix blocks $v_x$ can be performed independently, and the cost  scales linearly with respect to ${N_{\mathrm{f}}}$ (and therefore  $L$).  Once $v := \bigoplus_{x=1}^{N_\mathrm{f}} v_x $ is obtained, we may update the correlation potential as  
\begin{equation}
u\gets u+v=\bigoplus_{x=1}^{N_{\mathrm{f}}} (u_x+v_x).
\label{eqn:cp_update}
\end{equation}

Following the discussion of Section \ref{sec:convex}, we may readily formulate a convex optimization-based alternative to the least squares problem in Eq.~\eqref{eqn:lsq_local}. We may define the function 
$F^{\mathrm{act}}_x$ defined on the set of Hermitian $L_{A}\times L_{A}$
matrices by 
\[
F^{\mathrm{act}}_x(v_{x})=\mc{E}_{L_{A}}\left(\widetilde{f}_x+Ev_{x}E^{\top}\right). \quad 
\]
Note that we do not require $v_{x}$ to be traceless, since $v_{x}$ is only applied to the fragment instead of the entire impurity. Then if $0\prec D_{x}^{\mathrm{hl},\mathrm{frag}}\prec I_{L_{A}}$, the convex optimization problem 
\[
\inf_{v_{x}^{\dagger}=v_{x}}\left\{ \Tr[D_{x}^{\mathrm{hl},\mathrm{frag}}v_{x}]-F_{x}^{\mathrm{act}}(v)\right\} 
\]
 admits a solution $v_{x}^{\star}$. If $\widetilde{f}_{x}+v_{x}^{\star}$
has a gap between its $(L_{A})$-th and $(L_{A}+1)$-th eigenvalues
(ordered increasingly), then $\mathcal{D}(\widetilde{f}_{x}+Ev_{x}E^{\top},L_{A})$
has fragment block equal to $D_{x}^{\mathrm{hl},\mathrm{frag}}$,
i.e., we achieve exact fitting. If $\widetilde{f}_{x}+v_{x}^{\star}$
has no gap, then the ground state and 1-RDM are ill-defined, and, if the solution is unique, then 
there is no correlation potential $v$ that yields a well-defined 1-RDM
with exact fit. Furthermore, any optimizer $v^{\star}$ can be obtained from an optimizer 
$(v^\star, Z^\star, \alpha^\star)$ of the SDP 
\begin{equation}
\begin{aligned}
 & \underset{\stackrel{v_{x}\in\mathbb{C}^{L_{A}\times L_{A}},v_{x}^{\dag}=v_{x}}{Z\in\mathbb{C}^{(2L_{A})\times(2L_{A})}},\,\alpha\in\R}{\mathrm{minimize}} & \Tr(v_{x}D_{x}^{\mathrm{hl,frag}})-\alpha L_{A}+\Tr(Z)\\
 & \mathrm{subject\ to} & \widetilde{f}_{x}+ E v_{x} E^\top +Z-\alpha I\succeq0,\\
 &  & Z\succeq0.
\end{aligned}
\label{eqn:sdp_local}
\end{equation}

In the following discussion, the procedure above will be referred to as the local-fitting based DMET (L-DMET), which combines local correlation potential fitting and semidefinite programming. Note that DMET and L-DMET solve fixed-point problems of the same form \eqref{eqn:dmet_selfconsistency}, but with different choices of mappings $\mathfrak{F}$. We define the mappings associated with DMET and L-DMET as $\mathfrak{F}^{\text{DMET}}$ and $\mathfrak{F}^{\text{L-DMET}}$, respectively. As stated precisely in Proposition~\ref{prop:equiv} below, the fixed points of L-DMET and DMET are equivalent. Hence L-DMET introduces no loss of accuracy relative to DMET.
The proof is given in Appendix \ref{app:equivalency}.

\begin{prop}
\label{prop:equiv}
Suppose Eq. \eqref{eqn:dmet_selfconsistency} has a fixed point $u^{\star}$ with $\mathfrak{F}=\mathfrak{F}^{\text{DMET}}$, and $f+u^{\star}$ has a gap between
its $(N_{e})$-th and $(N_{e}+1)$-th eigenvalues (ordered increasingly). Let $D^{\mathrm{hl},\mathrm{frag}}\in \mc{S}$ be the associated high-level density matrix blocks, which satisfy $0\prec D_{x}^{\mathrm{hl},\mathrm{frag}}\prec I_{L_{A}}$
for $x=1,\ldots,N_{\mathrm{f}}$ and $\Tr[D^{\mathrm{hl},\mathrm{frag}}]=N_e$. Then $u^{\star}$ is a fixed point of Eq. \eqref{eqn:dmet_selfconsistency} with $\mathfrak{F}=\mathfrak{F}^{\text{L-DMET}}$. Similarly, under the same assumptions, if $u^{\star}$ if a fixed point of L-DMET, then it is also a fixed point of DMET. 
\end{prop}

In summary, L-DMET only leads to a modular modification of an existing DMET implementation. We provide a unified pseudocode for DMET and L-DMET  in Algorithm \ref{algo:unified}.  
\begin{algorithm}[H]
\caption{A unified pseudocode of DMET and L-DMET.}
\label{algo:unified}
\begin{algorithmic}[1]
        \Statex {\bf Input}: Initial low level density matrix $D^{\text{ll}, (0)}$, and chemical potential $\mu^{(0)}$.
        \Statex \quad \quad \quad \quad Partition the system  into $N_{\mathrm{f}}$ fragments.
        \Statex {\bf Output}: Correlation potential $u$ and high-level density matrix blocks $D^{\text{hl,frag}}$
        \While{correlation potential $u^{(k)}$ has not converged}
                \State Solve the ground state associated with $\hat H^\mathrm{ll}=\hat f+\hat c(u^{(k)})$ for $D^{\mathrm{ll},(k)}$
                \For{$x$ in $1,\dots, N_{\mathrm{f}}$}
                        \State Compute bath orbitals for impurity $x$.
                \EndFor
                \State Set $m=0,\nu^{(m)} = \mu^{(k)}$
                \While{chemical potential $\nu^{(m)}$ has not converged}
                        \For{$x$ in $1,\dots, N_{\mathrm{f}}$}
                                \State Solve the impurity problem $\hat{H}^{\mathrm{emb}}_{x}-\nu^{(m)}\hat N_x^\mathrm{frag}$ for $D_x^\mathrm{hl,frag}$.
                        \EndFor 
                        \State Use $\Tr(D^\mathrm{hl,frag})$ to update the chemical potential to  $\nu^{(m+1)}$.     
                        \State Set $m\gets m+1$.
                \EndWhile
                \State Set $\mu^{(k+1)} = \nu^{(m)}$.
                \If{DMET}
                        \State Update $u^{(k+1)}$ by solving the global correlation potential fitting problem.
                \EndIf 
                \If{L-DMET}
                        \State Update $u^{(k+1)}$ by solving the local correlation potential fitting problem.
                \EndIf
                \State Set $k\gets k+1$. 
        \EndWhile       
\end{algorithmic}
\end{algorithm}

\section{Other considerations}\label{sec:otherconsider}
  A related approach to improve the efficiency of the correlation potential fitting is called projected-based DMET (p-DMET) \cite{WuCuiTongEtAl2019}, which directly finds the closest low-level density matrix $D^\mathrm{ll}$ to the entire high-level density matrix $D^\mathrm{hl}$, subject to rank-$N_e$ constraints. This completely eliminates the correlation potential fitting procedure and is very efficient for large systems. It also eliminates the uncertainty introduced by the uniqueness of the correlation potential. However, it has also been observed that the result of the p-DMET has a stronger initial state dependence than DMET. In particular,when p-DMET
is used to study the phase diagrams of a 2D Hubbard model, the resulting phase boundary from p-DMET is blurrier than that obtained from DMET \cite{WuCuiTongEtAl2019}. On the other hand, Proposition \ref{prop:equiv} guarantees that the fixed points of L-DMET and DMET are the same. We will also demonstrate by numerical results that L-DMET and DMET can produce identical phase diagrams. 

When the two-body interaction term $(pr|qs)$ is nonlocal (such as in the case of quantum chemistry calculations), one often replaces $\hat f$ in Eq. \eqref{eqn:H_ll} by $\hat f=\hat{f}(D^{\text{ll}})$, which is a Fock operator that depends on the low-level density matrix $D^{\text{ll}}$. Then  Eq. \eqref{eqn:H_ll} needs to be solved self-consistently as in the case of solving Hartree-Fock equations. Such an extra self-consistency step at the low level is also called {\it charge self-consistency} \cite{cui2019efficient} and can be used to take into account long-range interactions beyond the sparsity pattern of $\mc{S}$.

When $f+u^{\star}$ is gapless, the corresponding low-level density matrix $D^{\text{ll}}$ is ill-defined (even though $u^{\star}$ itself may still be well-defined via the semidefinite programming formulation of correlation potential fitting),  and the self-consistent iteration of Eq. \eqref{eqn:dmet_selfconsistency} cannot proceed without modification. One possibility is to use the recently developed finite temperature DMET \cite{sun2019finite}. The other possibility is to generate a mixed=state low-level density matrix using a Fermi-Dirac smearing with a low temperature, and extract the bath orbitals from the density matrix directly (see Appendix \ref{sec:bath}). We remark that both options formally violate the original premise of DMET, namely the Schmidt decomposition of a Slater determinant \cite{DMET2012,DMET2013} or the CS decomposition as in Eq. \eqref{eqn:c_ll}. A proper treatment of gapless systems remains a future research direction.

\section{Numerical experiments: 2D Hubbard model}
\label{sec:2d_hubbard}

The 2D Hubbard model can describe rich physical phenomena including phase transitions \cite{Hideo2004}, superconductivity \cite{boxiao2016}, charge and spin density waves \cite{Tewordt1997,Fujita1990}, stripe order \cite{zheng2017stripe,Mizusaki:2006jd}, etc. Here we report the performance of L-DMET for the 2D Hubbard model on a square lattice with  periodic boundary conditions. 

The fragment size is set to $2\times 2$. The initial guess and the low-level density matrix are generated by the unrestricted Hartree Fock (UHF) method. When the system becomes gapless, we use Fermi-Dirac smearing with $\beta=100$ (i.e., temperature $T=0.01$ in the unit of the hopping parameter $t=1$) according to the discussion in Section \ref{sec:otherconsider}. The finite temperature smearing in zero-temperature DMET is a numerical regularization technique. Smaller choices for $\beta$ correspond to more severe regularization and reduced accuracy in the solution of DMET. 
In fact, in order to improve numerical convergence in of the least squares fitting procedure \eqref{eqn:least_squares}, we \emph{always} to add a temperature (always set to $T = 0.01$) within the fitting procedure itself. Hence we in fact solve \eqref{eqn:least_squares} where the map $\mathcal{D}$ is understood to indicate the appropriate density matrix at temperature $T=0.01$. The bath orbitals are then extracted from the resulting finite-temperature density matrix via the same approach as described above.

The impurity problems are solved by full configuration interaction (FCI) implemented in \texttt{PySCF}. The number of orbitals in each impurity problem is fixed to be $8$ orbitals. We present results for both DMET and L-DMET. Within DMET we solve the least squares problem \eqref{eqn:least_squares} using BFGS via \texttt{SciPy}, and within L-DMET we solve the SDP \eqref{eqn:sdp_local} with a splitting conic solver (SCS)\cite{ocpb:16,scs} called via \texttt{CVXPY} \cite{cvxpy,cvxpy_rewriting}. For both of the methods, the convergence tolerance is set to be $10^{-8}$. The convergence criterion of the DMET and L-DMET fixed point problem is set to
\begin{equation}
  \label{eq:8}
  \frac{|E^{(k)}-E^{(k+1)}|}{|E^{(k)}|} < 10^{-8} \quad \text{ and } \quad \frac{\|D^{\text{hl}, (k)}-D^{\text{hl}, (k+1)}\|_F}{\|D^{\text{hl}, (k)}\|_F}<10^{-6}.
\end{equation}

\subsection{Comparison of semidefinite programming  and least squares fitting}
\label{sec:SDPcompare}
Before presenting an overall comparison of DMET and L-DMET, we first present a comparison of the two approaches to the global correlation potential fitting procedure presented above, namely the least squares approach \eqref{eqn:least_squares} (interpreted at finite temperature $T=0.01$ to improve numerical convergence, as discussed above) and the SDP approach \eqref{eqn:min_sdp}. Our results in this section compare these two approaches for the \emph{first} correlation potential fitting step of DMET, initialized from UHF on the $6\times 6$ 2D Hubbard model.

We measure the success rates of the two methods as follows. For a given on-site interaction strength $U$ and filling factor $n$ (i.e., the number of electrons divided by the number of sites), the success rate is defined as 
\begin{equation}
\text{success rate} = \frac{\text{number of successful samples}}{\text{number of total samples}}
\end{equation}
Each sample is specified by a random potential (each entry of which is sampled independently from the uniform distribution $\mc{U}[-0.1,0.1]$), which is added to the one-body Hamiltonian. The total number of samples is $1000$. The least squares method fails if the norm of the gradient is greater than $10^{-8}$ after 2000 iterations. The SDP method fails if any of the primal residual, the dual residual and the duality gap is greater than $10^{-9}$ after $2500$ steps. The success rate is measured for multiple values of both $U$ and $n$.

\begin{figure}[!htp]
        \includegraphics[scale=0.5]{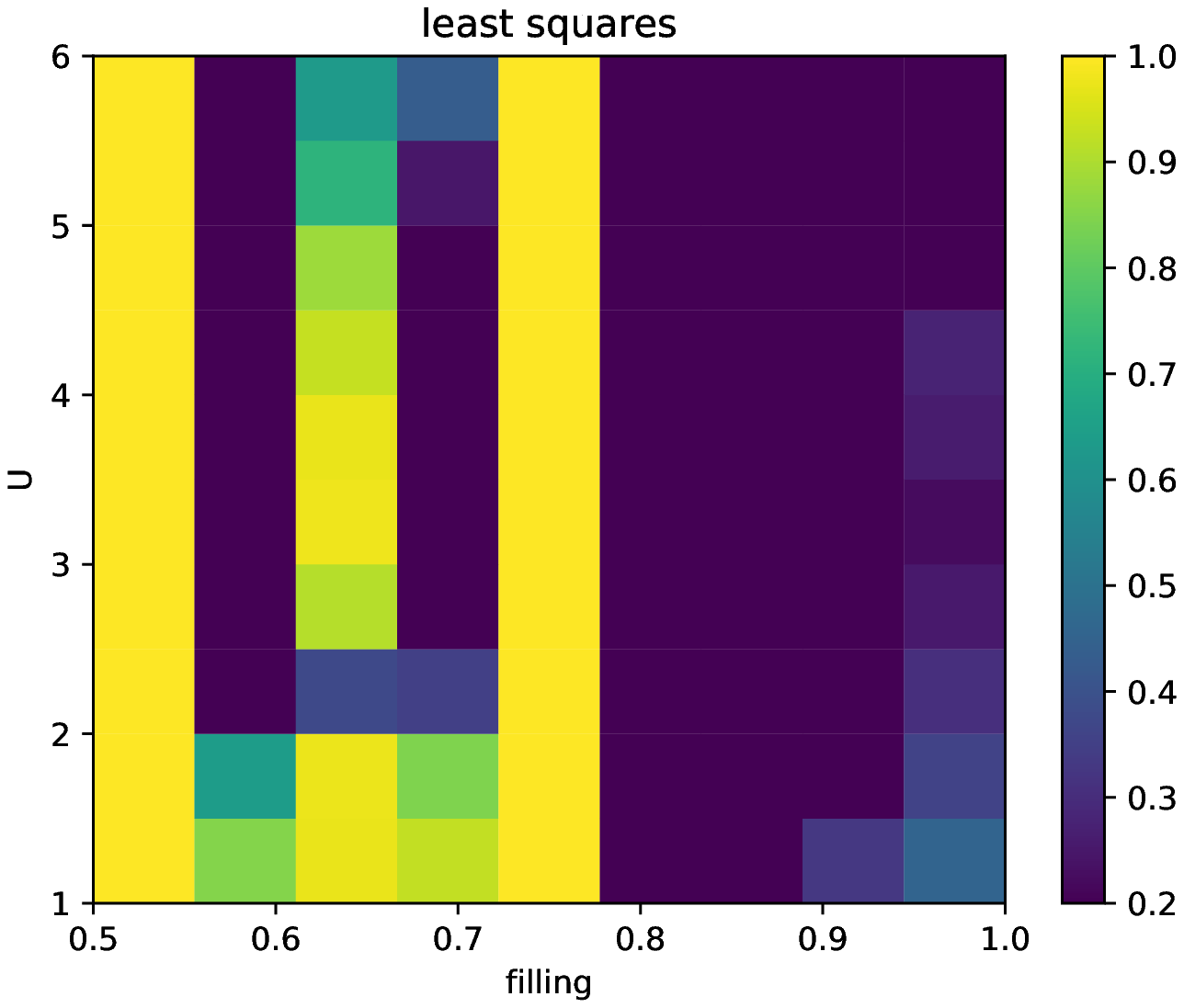}
        \includegraphics[scale=0.5]{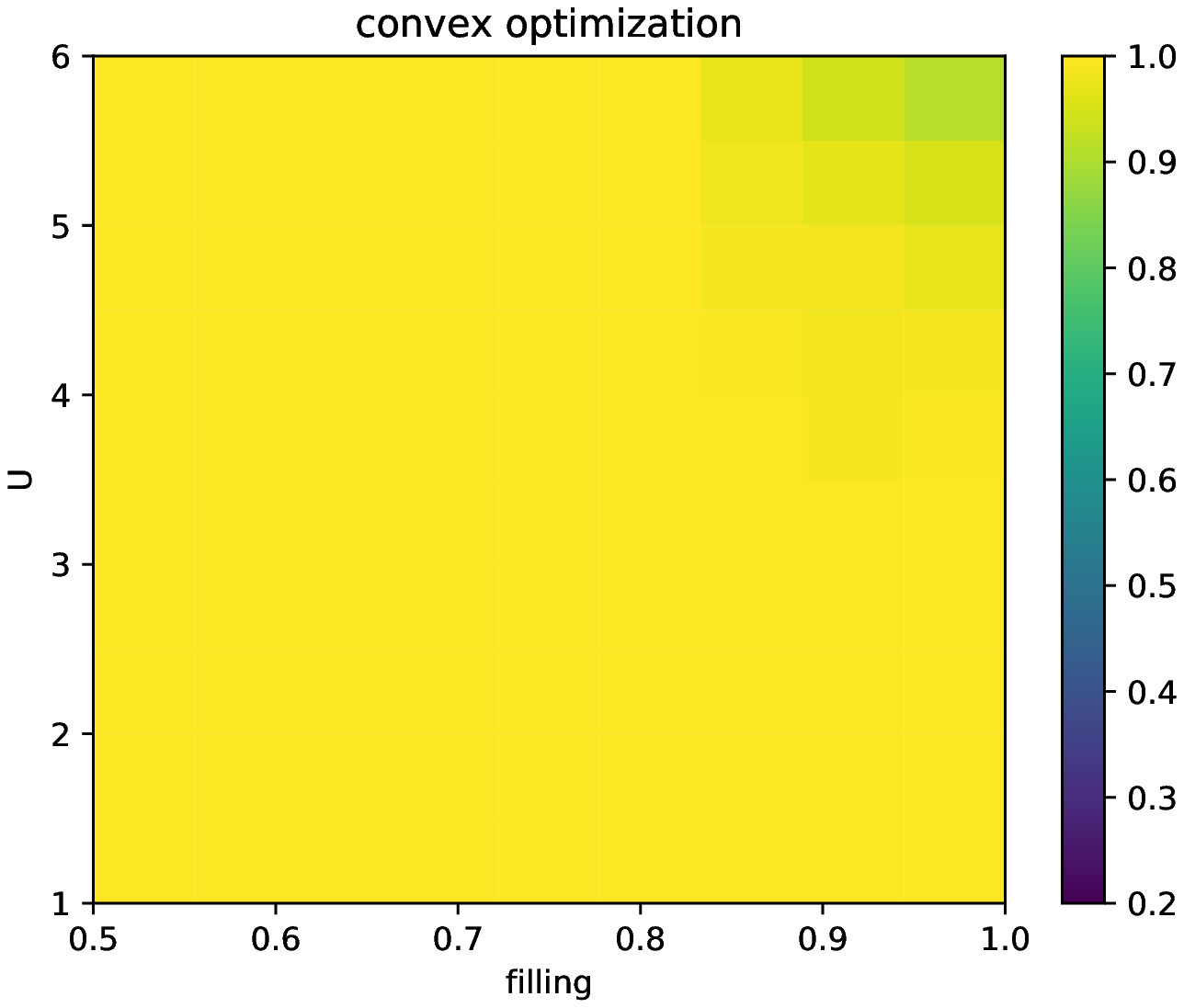}
        \caption{Success rates of the least squares (left) and convex optimization (right) approaches for $1000$ samples of a $6\times 6$ 2D Hubbard model with random potential.}
        \label{fig:robust}
\end{figure}

Fig. \ref{fig:robust} shows that semidefinite programming 
is much more robust than the least squares approach, despite the fact that the least squares fitting is performed with some finite temperature smearing. 
The least squares procedure can reliably converge only when the number of electrons is $18$ and $26$. Typically, the least squares approach is robust at half-filling without a random potential. However, when the random potential is added, the least squares approach fails frequently. On the other hand, the SDP method succeeds consistently across most test cases. The lowest success rate of the SDP method (around $90\%$) occurs near U = 6.0 at half-filling.
The success rate is nearly $100\%$ in all other cases.

We summarize the results for the correlation potential fitting as follows: as a regularization technique, the finite temperature smearing can improve the robustness of the least squares approach in the gapless case. However, the regularized problem may still be ill-conditioned to solve using solvers such as BFGS.  On the other hand, the SDP approach is parameter-free. The numerical tests show that the SDP reformulation significantly increases the robustness of the correlation potential fitting.

We also present the comparison between the performance of SDP and least squares fitting for the 1D Hubbard model, where we observe that the success rate of SDP is $100\%$. These results are reported in Appendix \ref{sec:comp_hubbard_1d}.

\subsection{Phase diagram}

For strongly-correlated systems, single-particle theories such as RHF/UHF may produce qualitatively incorrect order parameters, leading to incorrect phase diagrams. We expect that DMET/L-DMET can correct order parameters through the self-consistent iteration for the correlation potential. Without self-consistent iteration, the phase boundary of DMET/L-DMET tends to be very similar to that of UHF. As an example, we study the phase transition between antiferromagnetism and paramagnetism as studied in \cite{WuCuiTongEtAl2019}. We impose the constraint that all impurities should be translation-invariant (TI). The TI constraint is crucial for improving the convergence behavior of DMET and L-DMET especially around the phase boundary. 

\begin{figure}[!htp]
  \centering
  \includegraphics[scale=0.5]{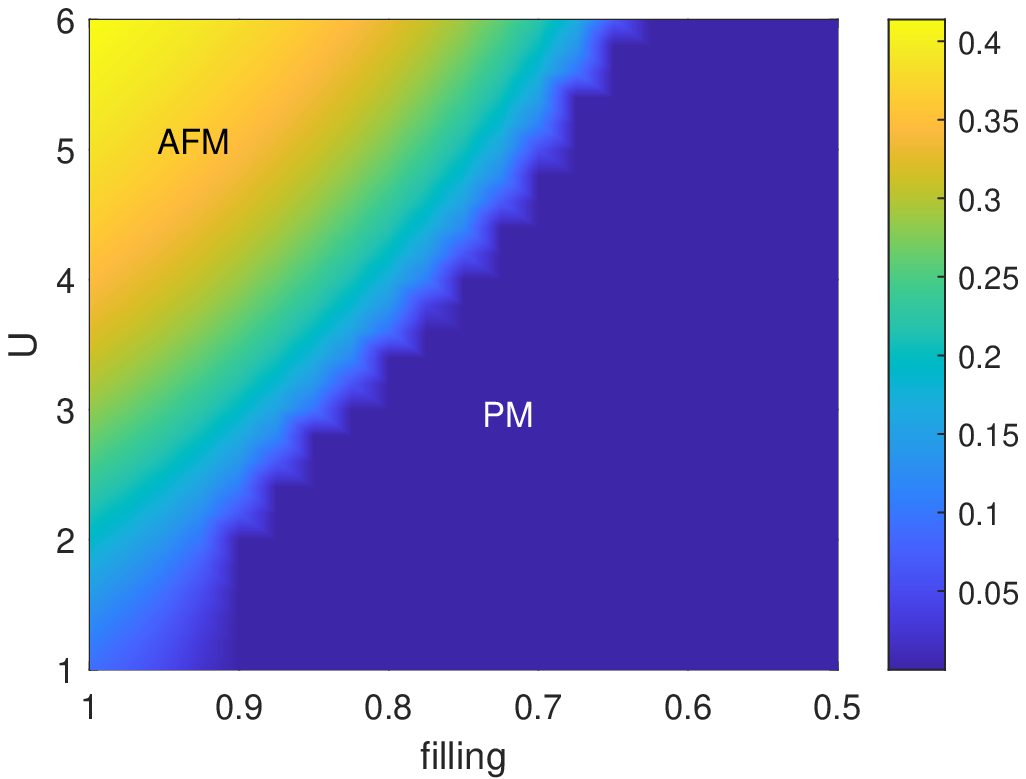}
  \includegraphics[scale=0.5]{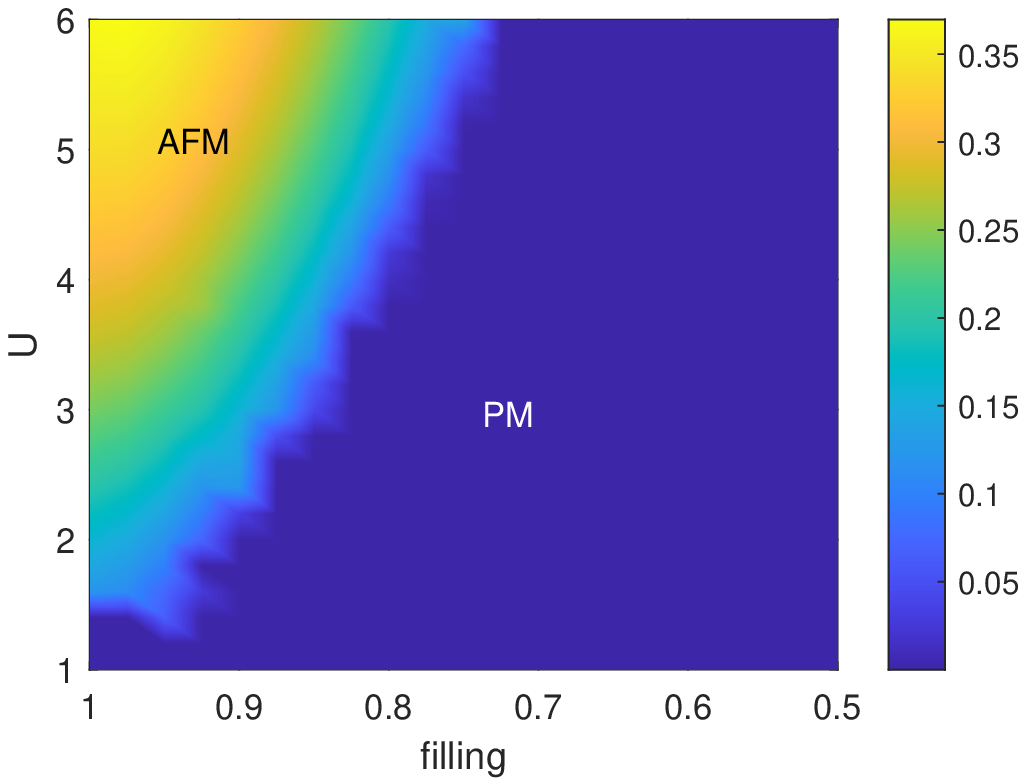}
  \includegraphics[scale=0.5]{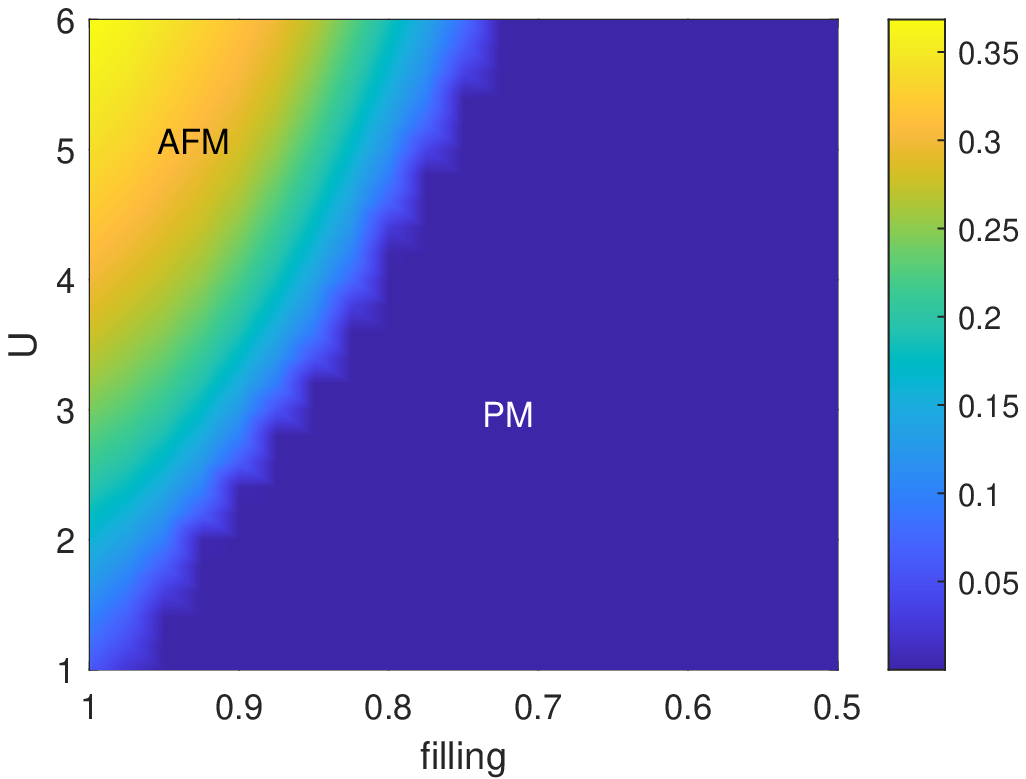}
  \caption{The comparison of phase diagrams of UHF (left), DMET (middle) and L-DMET (right) for the 2D Hubbard model. The color indicates the spin polarization. 
  }
  \label{fig:phase_diagram}
\end{figure}

We perform a series of computations on a $20 \times 20$ lattice. The fragment size is $2\times 2$. 
The filling factor $n$ and the interaction strength $U$ define two axes of the phase diagram. We consider $21$ uniformly-spaced values of of $n$ and $26$ uniformly-spaced values of $U$. We use the spin polarization to identify the phases. The spin polarization is defined as 
$$m=\frac{|\Tr(D^{\text{hl},\uparrow})-\Tr(D^{\text{hl},\downarrow})|}{\Tr(D^{\text{hl},\uparrow}) + \Tr(D^{\text{hl},\downarrow})}.$$
$D^{\text{hl},\uparrow}$ and $D^{\text{hl},\downarrow}$ are, respectively, the spin-up and spin-down components of the high-level global density matrices. The spin polarization as a function of $n$ and $U$ is presented in Fig. \ref{fig:phase_diagram}. The phase diagrams of DMET, and L-DMET are almost identical except for certain points on the phase boundary. Both are significantly  different from the UHF phase diagram. The lower-left corner of the phase diagram corresponds to gapless low-level systems, and the phase diagrams obtained from DMET and L-DMET slightly differ here. The performance of L-DMET is better than the previously proposed p-DMET method \cite{WuCuiTongEtAl2019}, which leads to a slightly blurred phase boundary.





\subsection{Robustness with respect to the initial guess}
\label{sec:robustness}

We now demonstrate the numerical stability of the L-DMET method with respect to the initial guess. We consider two filling factors for the $6\times 6$ 2D Hubbard model at $U=4$: filling $n=1.0$ ($36$ electrons) and filling $n=0.5$ ($18$ electrons), for which the solution is in the  antiferromagnetic (AFM) and paramagnetic (PM) phase, respectively. For all calculations in this section, we take the fragment size to be $2 \times 2$. 

To show that L-DMET is also effective when the system is inhomogeneous, we explicitly break the translation symmetry by  introducing a random on-site potential. 
Each entry of the random potential is sampled independently from the uniform distribution $\mc{U}[-0.2,0.2]$.  We deliberately choose the initial guess to have the wrong order parameter in order to test the robustness of the algorithm. We choose initial guesses for the DMET loop by incompletely converging the self-consistent field iteration for UHF (i.e., terminating after a fixed number of iterations). We in turn initialize our UHF calculations with hand-picked initial guesses; since the self-consistent iteration for UHF is terminated before convergence, the result (which we use as our initialization for DMET) depends on the initial guess.

In the AFM case ($n=1.0$), the initial guess for UHF is chosen to be a state in the PM phase, which is obtained by alternatively adding/subtracting a small number ($10^{-3}$) to the uniform density according to a checkerboard pattern. In the PM case, we initialize UHF in the AFM phase with spin-up and spin-down densities of $0.1$ and $0.4$ respectively.
We terminate UHF after the $1$st, $5$th, and $10$th iterations to provide initial guesses for DMET and L-DMET.

\begin{figure}[htbp]
  \centering
  \includegraphics[scale=0.6]{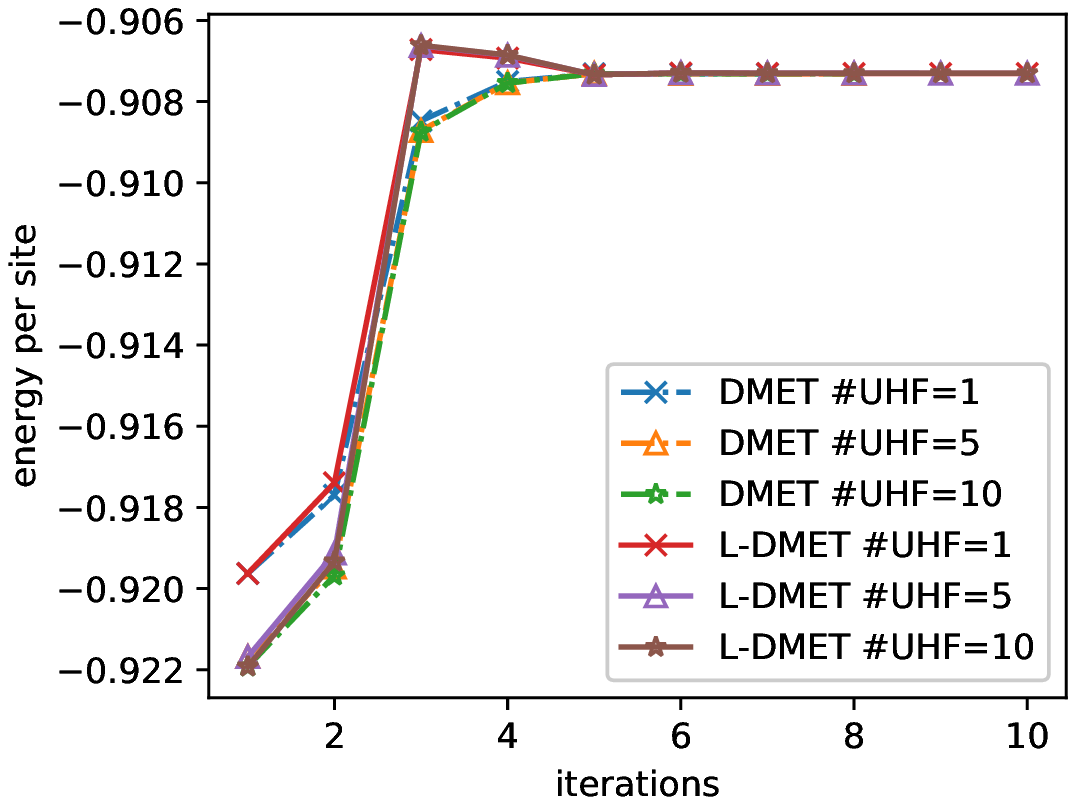}
  \includegraphics[scale=0.6]{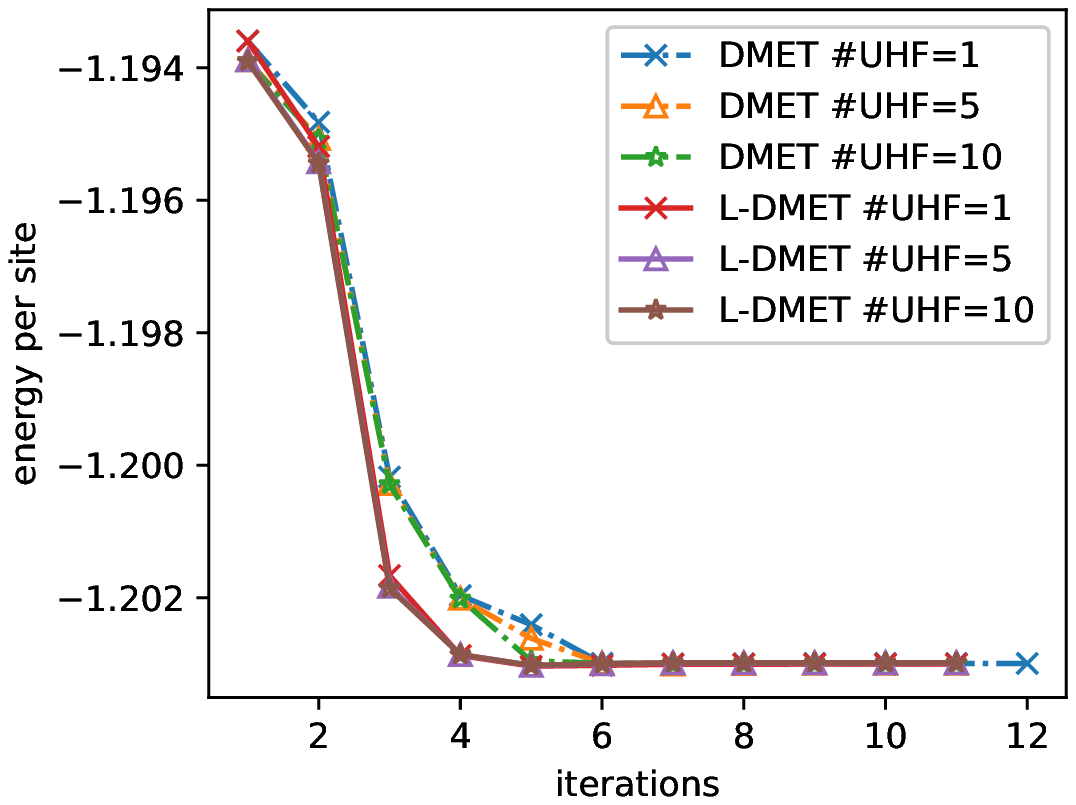}\\
  \caption{The convergence of the energy per site for DMET and L-DMET with different initial guesses. Results are shown for the$6 \times 6$ Hubbard model at $U = 4.0$ with fragment size $2\times 2$. The filling is specified by $n=0.5$, $18$ electrons (left) and $n=1.0$, $36$ electrons (right). The initialization procedure for the figures at left and right are described in the text of section \ref{sec:robustness}.} 
  \label{fig:initdepend}
\end{figure}

For both DMET and L-DMET, we use DIIS to accelerate the convergence starting from the second iteration. We compare the convergence of DMET and L-DMET with the same random potential in Fig. \ref{fig:initdepend}. Both
DMET and L-DMET converge to the same fixed point within $12$ iterations
with different initial guesses. This experiment verifies two crucial
features of L-DMET: (1) L-DMET reaches the same solution as DMET at self-consistency, when the low-level model is gapped, and (2) in both the PM phase and AFM phase, the fixed point of L-DMET is independent of the choice of the initial guess. More specially, with different unconverged UHF initial guesses, L-DMET always converges to the same fixed point as DMET does.



\subsection{Jacobian of the fixed point mapping}

\cref{fig:initdepend} shows that the number of iterations needed for L-DMET to converge is approximately the same as for DMET, starting from a range of initial guesses. The same behavior is also observed for all the numerical tests presented in this paper. This finding is somewhat counterintuitive, given that L-DMET updates the correlation potential only locally, while DMET can use the information of the global density matrix and update the correlation potential globally. From the perspective of solving the fixed point problem in \cref{eqn:dmet_selfconsistency}, the convergence rate in the linear response regime is largely affected by the properties of the Jacobian matrix of $\mathfrak{F}\circ\mathfrak{D}$, where the mapping $\mathfrak{F}$ stands for  $\mathfrak{F}^{\text{DMET}}$ and $\mathfrak{F}^{\text{L-DMET}}$ in DMET and L-DMET, respectively.  

\begin{figure}[!htb]
  \centering
  \includegraphics[scale=0.55]{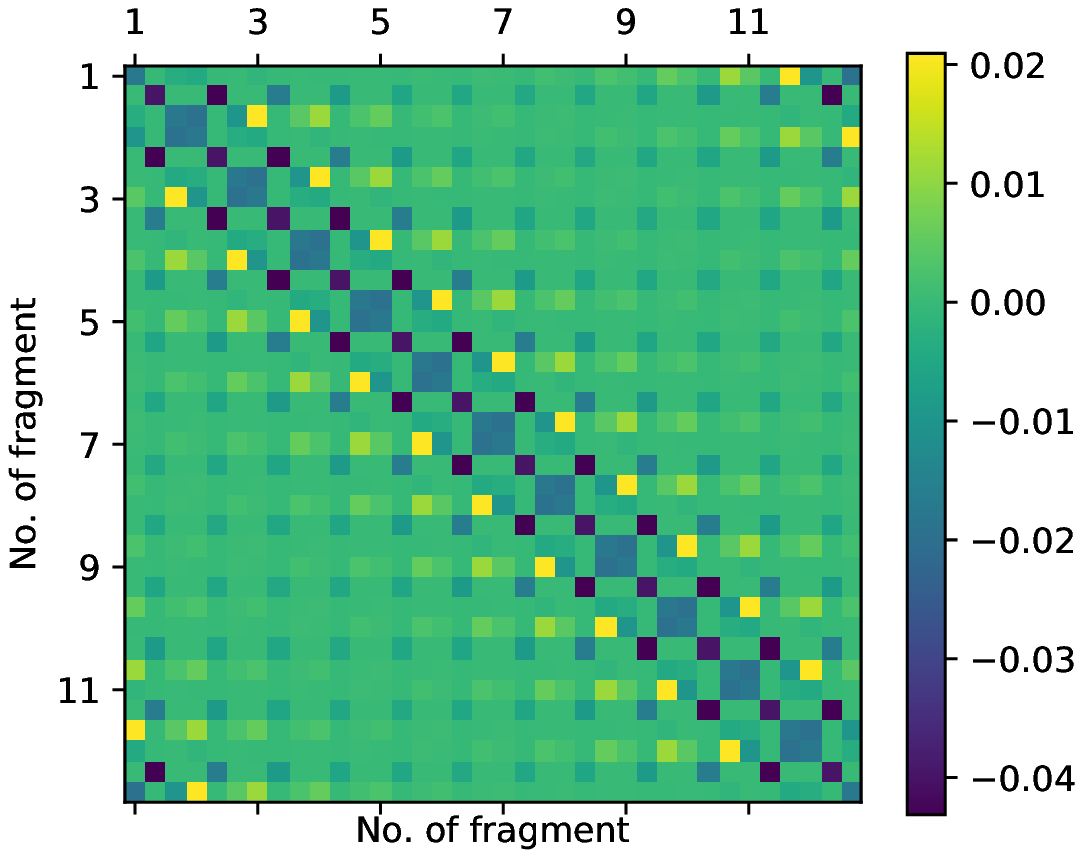}
  \includegraphics[scale=0.56]{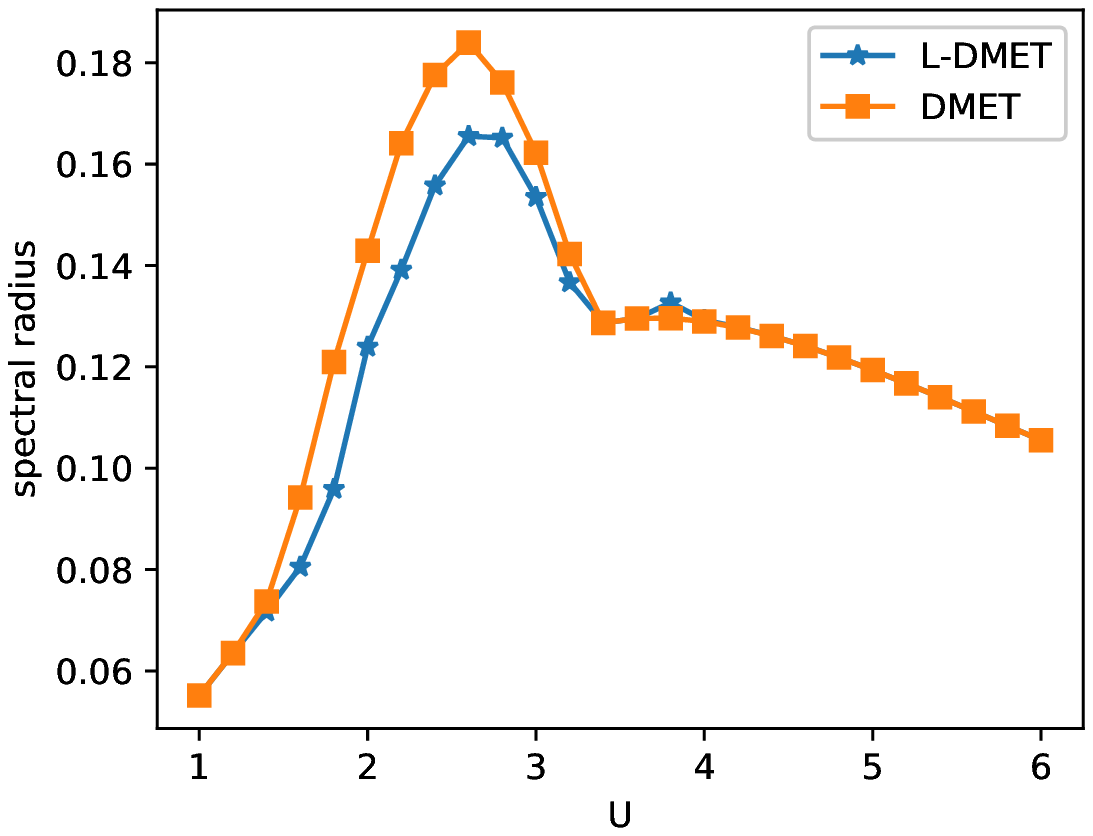}
  \caption{The matrix ${\partial D^\text{hl,frag}}/{\partial u}$ (left) at $U=4.0$, and the spectral radius of the Jacobian of Eq. \eqref{eqn:dmet_selfconsistency} for DMET and L-DMET (right). There are 3 degrees of freedom within $u_x$ and $D^\text{hl,frag}_x$ for each fragment $x$; hence the size of the matrix ${\partial D^\text{hl,frag}}/{\partial u}$ is $36\times 36$, viewed as a $12\times 12$ matrix of $3 \times 3$ blocks.}
  \label{fig:locality}
\end{figure}

To illustrate the properties of the Jacobian, we consider 1D Hubbard model with $24$ sites with anti-periodic boundary condition. The total number of electrons is $24$ (i.e., half-filling). Each fragment has 2 sites. The low-level method is the restricted Hartree-Fock method. We investigate two quantities in the self-consistent equation in Eq. \eqref{eqn:dmet_selfconsistency}: the linear response of $D^\text{hl,frag}_x$ with respect to $u$, i.e., $R = {\partial D^\text{hl,frag}_x}/{\partial u}$ and the Jacobian matrix of Eq. \eqref{eqn:dmet_selfconsistency}. As shown in  \cref{fig:locality} (a), the matrix $R$ is highly localized. This means that the response of the density matrix block $D^\text{hl,frag}_x$ is relatively small with respect to the perturbation of the  correlation potential $u_y$ in another fragment $y$, when $x$ and $y$ are far apart from one other. Such `near-sighted' dependence implies that the local update procedure can also lead to an effective iteration scheme.  

\cref{fig:locality} (b) shows that the spectral radius of the Jacobian matrix of DMET and that of L-DMET are relatively small.  For the range of $U$'s studied, the spectral radius is uniformly smaller than $0.2$ (and in particular smaller than $1$). Hence $\mathfrak{F}\circ\mathfrak{D}$ can define a contraction mapping even without mixing, and as such the fixed point problem in \cref{eqn:dmet_selfconsistency} is easy to solve. It is also interesting to observe that the spectral radius peaks around $U\approx 2.5$. For larger value of $U$, the spectral radius decreases with respect to $U$, indicating that the DMET/L-DMET iterations are easier to converge numerically.

\subsection{Efficiency}
\begin{figure}[!htp]
        \centering
        \includegraphics[scale=0.6]{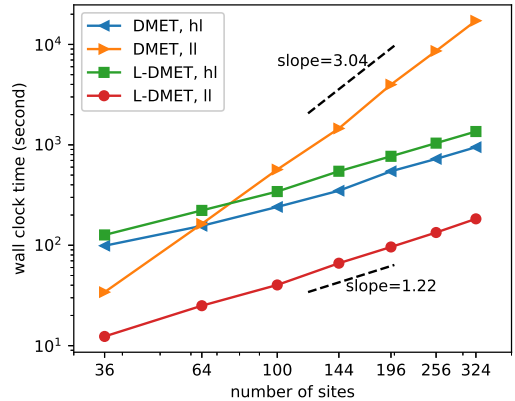}
        \caption{Computational cost of DMET and L-DMET calculations for 2D Hubbard models. The CPU time is averaged over 20 experiments.}
        \label{fig:cputime_hubbard}
\end{figure}

L-DMET mainly reduces the computational cost at the single-particle level (i.e., low level). The CPU time for 2D Hubbard systems ranging from size  $6\times 6$ to $18\times 18$ (with fragment size $2\times 2$ in all cases) is reported in  Fig. \ref{fig:cputime_hubbard}.
We report the time of the low-level and high-level calculations separately. Each calculation is performed on a single core.  The cost of the low-level calculations in DMET grows as $\Or(L^{3.04})$, while the cost of low-level calculation in L-DMET is reduced to $\Or(L^{1.22})$. When the number of sites is $324$, L-DMET is $49$ times faster than DMET for the low-level calculations.

\section{Numerical experiments: Hydrogen chain}
\label{sec:hydrogen}

\subsection{Efficiency and accuracy}

In this section, we consider the application of L-DMET to a real quantum-chemical system, the hydrogen chain. The Hamiltonian is discretized using the STO-6G basis set, and these basis functions are orthogonalized with the L\"owdin orthogonalization procedure. We use open boundary conditions and the restricted Hartree-Fock (RHF) method for the low-level method. The  chain is partitioned into fragments with $2$ adjacent atoms in each fragment, and the fragments do not overlap with each other. The high-level problem is solved with the full configuration-interaction (FCI) method. The CPU times for the low-level and the high-level parts of the calculation are aggregated separately over the entire self-consistent loop.

\begin{figure}[!htp]
  \centering
  \includegraphics[scale=0.5]{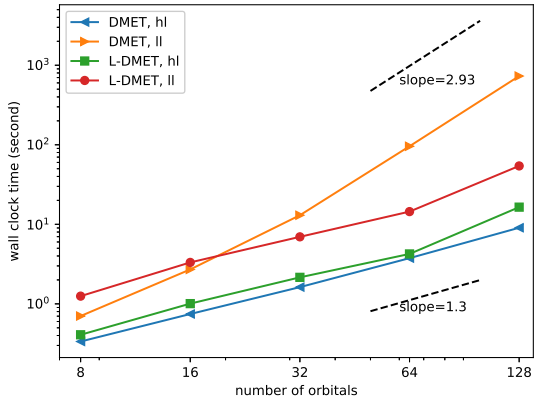}
  \caption{Computational cost of DMET and L-DMET calculations for hydrogen chains.}
  \label{fig:h10_cputime}
\end{figure}

\cref{fig:h10_cputime} shows that as the system size increases, the costs of both DMET
and L-DMET calculations are dominated by the low-level calculations. As a result, L-DMET is significantly faster than DMET due to the acceleration of the low-level calculations.  When the number of orbitals is 128 (i.e., 128 atoms), the low-level part of L-DMET is $13.5$ times faster than that of DMET.  Meanwhile, the wall clock times for the high-level parts of DMET and L-DMET are comparable for all systems considered. 

\begin{figure}[!htp]
  \centering
  \includegraphics[scale=0.5]{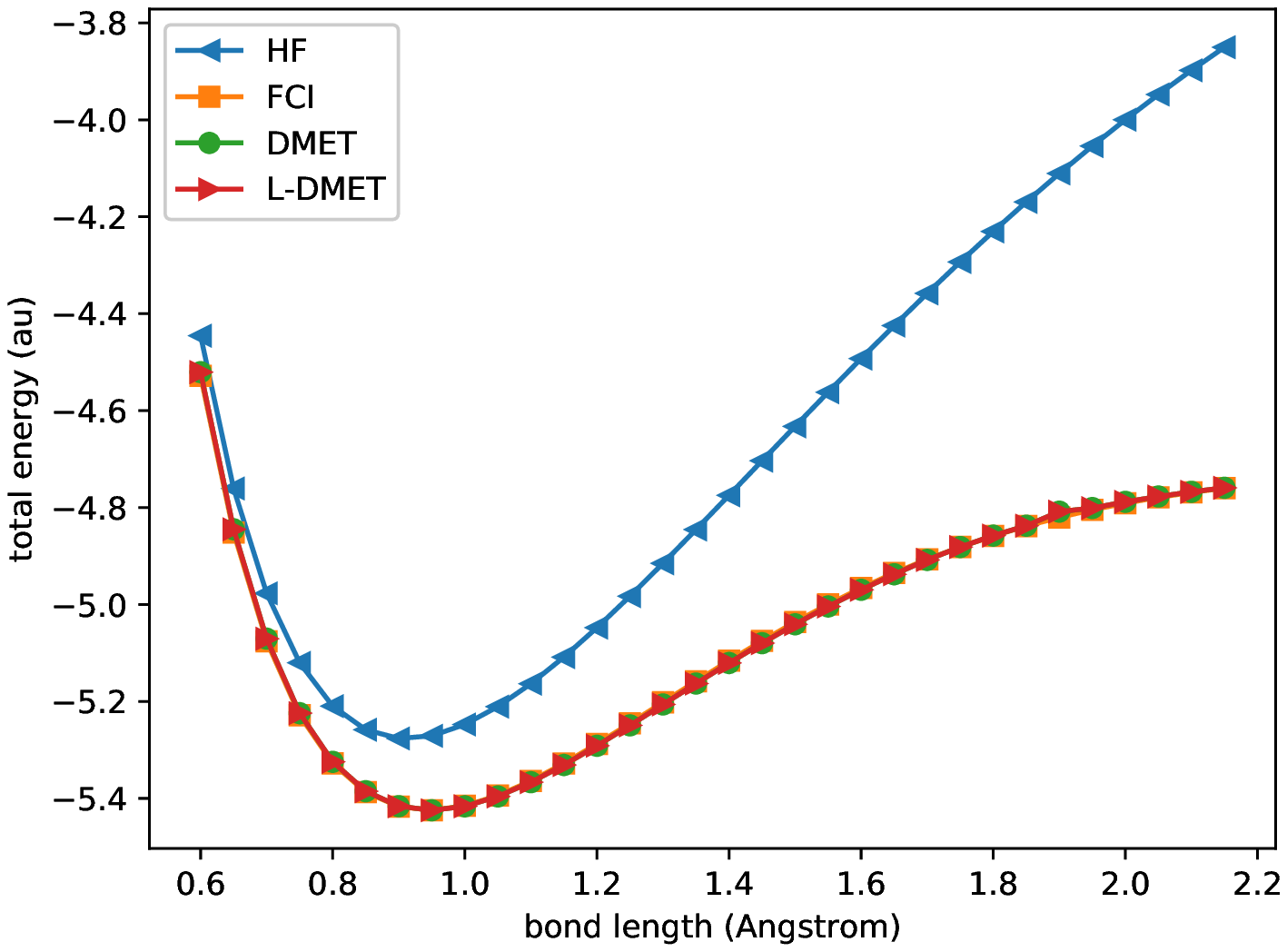}
  \includegraphics[scale=0.5]{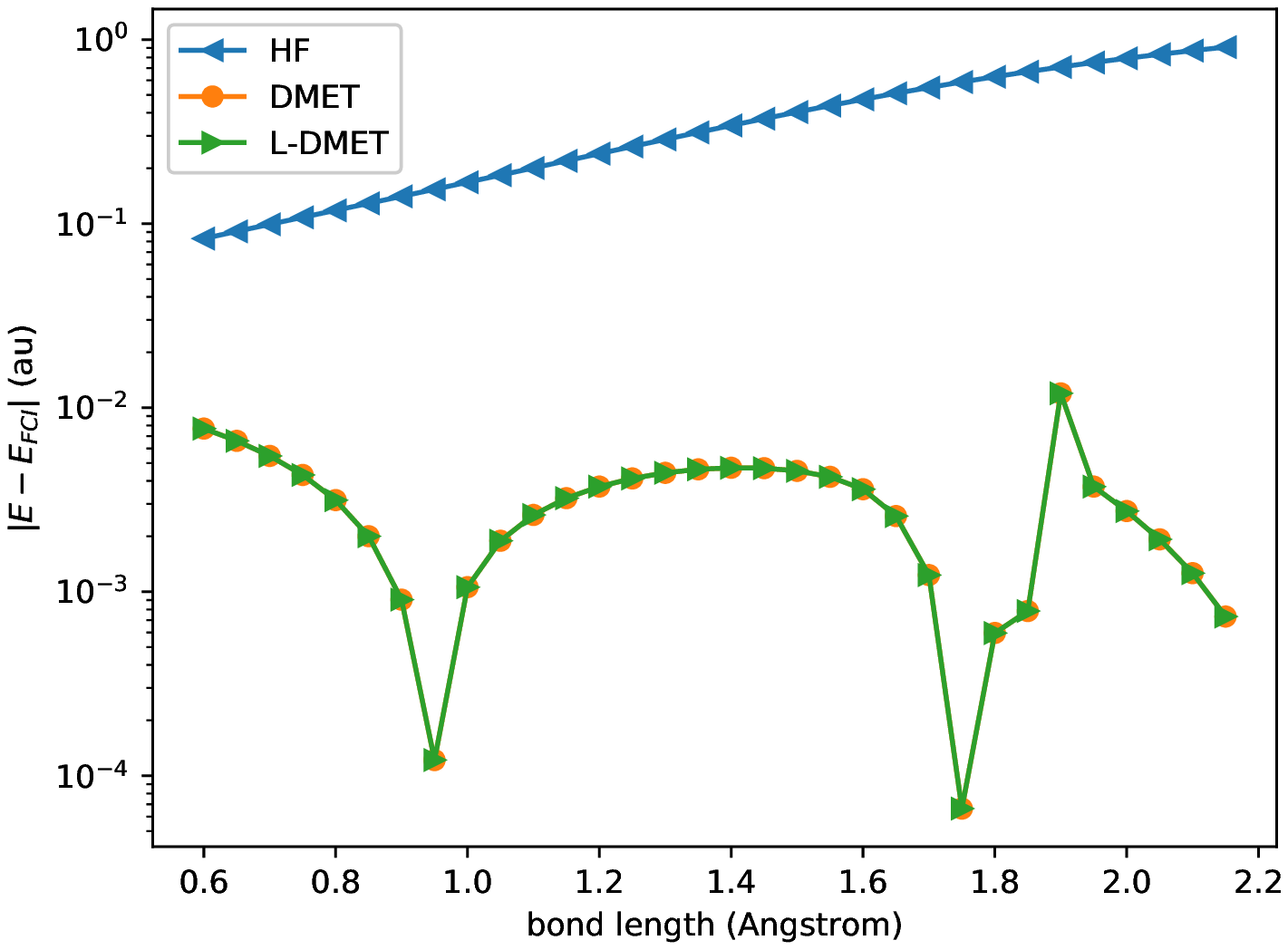}
  \caption{Total energy of the  hydrogen chain as computed by HF, FCI, DMET, and L-DMET (left). Error of the total energy as computed by HF, DMET, and L-DMET (right).}
  \label{fig:hchain}
\end{figure}

To demonstrate the accuracy of L-DMET, we report the dissociation energy curve  for a hydrogen chain with $10$ atoms. We start from an equidistant configuration and stretch the hydrogen chain, maintaining equal distances between atoms. The total energy curves of RHF, FCI, DMET, and L-DMET are shown in Figure \ref{fig:hchain}. The DMET and L-DMET curves are indistinguishable at all bond lengths. Compared to the exact value (FCI energy), the total energy errors of DMET and L-DMET are uniformly less than $0.01$ a.u.

\subsection{Impact of the fragment size}
To improve the accuracy of DMET calculations, one may consider increasing the fragment size.
However, we demonstrate that larger fragments can lead to numerical difficulties. In particular, we observe that a large fragment size can easily lead result in a gapless low-level model, which complicates the self-consistent iterations in DMET/L-DMET calculations.

We demonstrate the issue using a hydrogen chain with $36$ atoms and a bond length of 1 a.u. The coupled cluster method with singles and doubles (CCSD) is employed to solve impurity problems. 
 We consider different partitions of the orbitals specified by fragment sizes of $1$, $2$, $3$, $4$, $6$, $9$, and $12$. Furthermore, we experiment with multiple fitting strategies 
 for each partitioning of the entire system by performing correlation potential fitting using possibly finer partitions of the system. For instance, when the fragment size is $6$ and there are $4$ fragments, we may choose to perform correlation potential fitting by considering only the diagonal blocks of size $3$, so that there are $8$ blocks in total. This second block size will be referred to the `fitting size,' as opposed to the `fragment size' which specifies the size of the impurity problems that are solved. Note that the fragment size must be a multiple of the fitting size.  When the fitting size is 1, DMET reduces to the density embedding theory (DET)\cite{bulik2014density}.
According to \cref{app:unique}, when the fitting size is too large, the correlation potential may not be unique.  

\begin{figure}[!htp]
  \centering
  \includegraphics[scale=0.5]{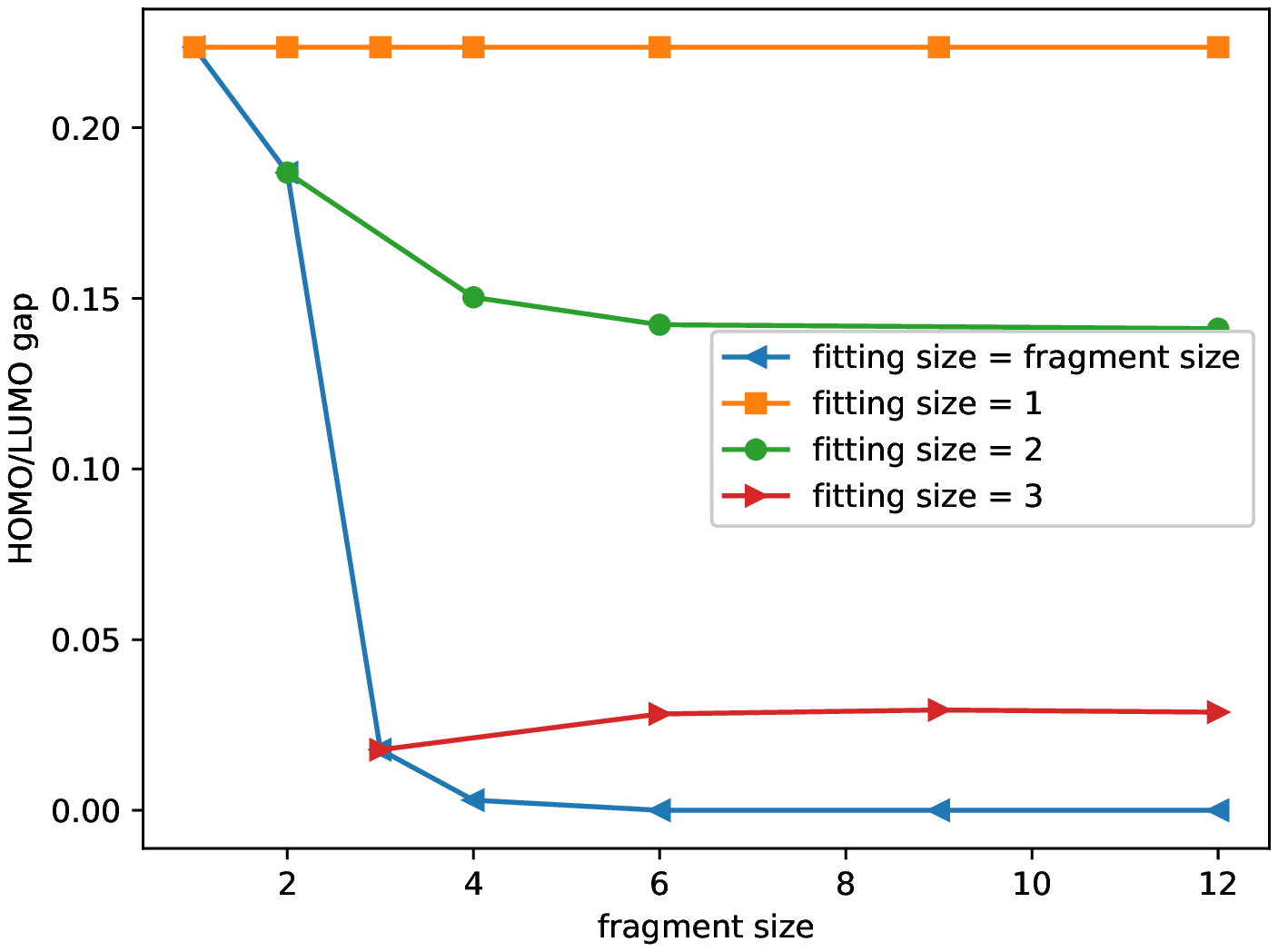}
  \includegraphics[scale=0.5]{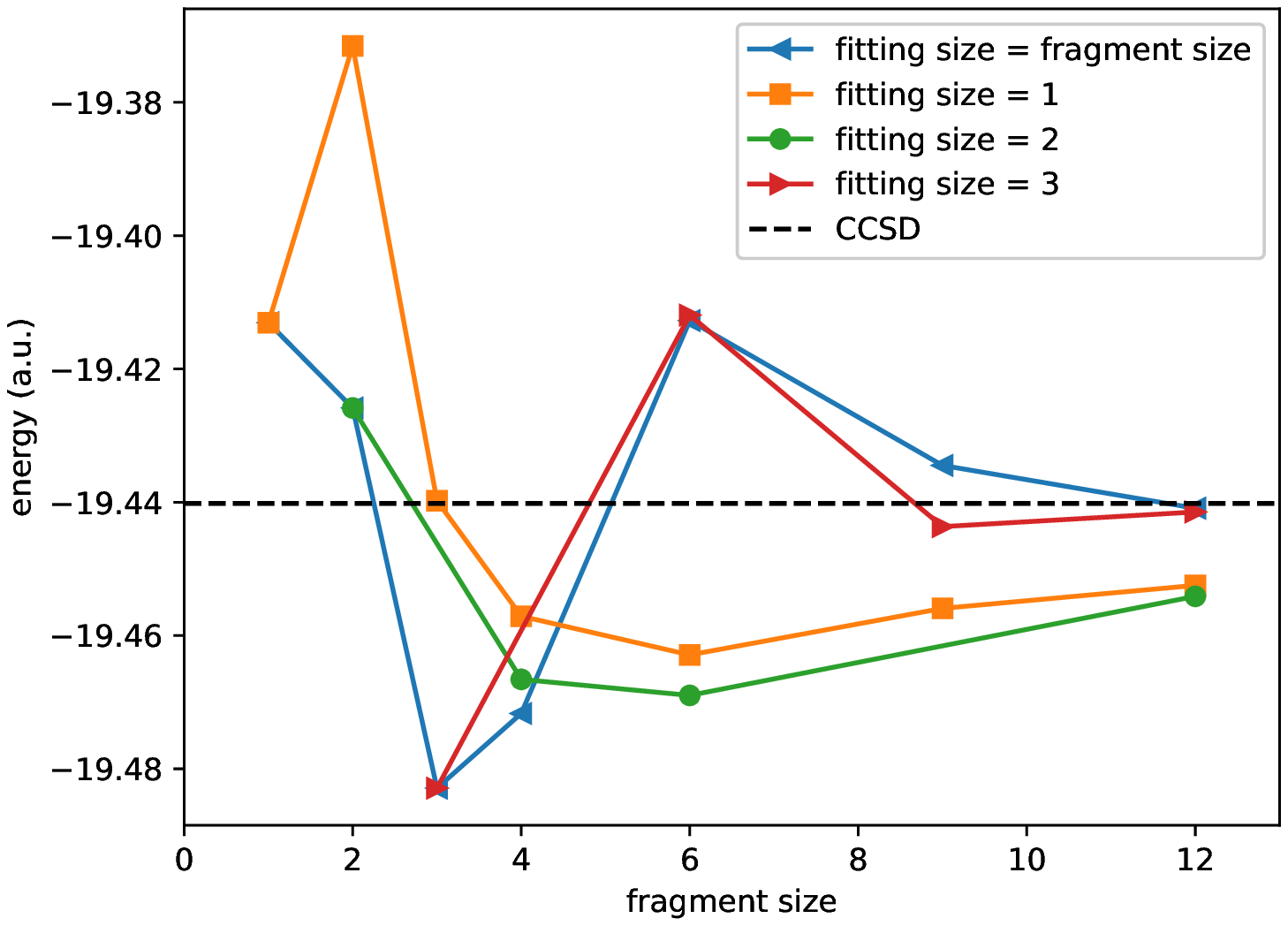}
  \caption{Low-level HOMO-LUMO gap (left) and total DMET energy (right) for different fragment sizes and fitting sizes. The dotted line (right) shows the CCSD energy for the entire system as a reference.}
  \label{fig:gap}
\end{figure}

As shown in Figure \ref{fig:gap}, when the fitting size is set to be the same as the fragment size in DMET, the gap of the low-level Hamiltonian decreases as the fragment size increases. It eventually vanishes when the fragment size is greater than $6$. However, if we fix the fitting size, 
the gap tends to be a constant as the fragment size increases. The same observations apply for L-DMET. 
Different fitting sizes also lead to different convergence patterns of the total energy as the fragment size increases. After enough iterations, the total energies computed with different fitting sizes become comparable, but we observe that the DMET and L-DMET self-consistent iterations are more stable when the low-level energy gap does not vanish. 


\section{Conclusion}\label{sec:conclusion}

In this paper, we propose the L-DMET method for tackling the problem of correlation potential fitting in the density matrix embedding theory (DMET). This is often a computational bottleneck in large-scale DMET calculations, particularly for inhomogeneous systems. L-DMET improves the robustness of the correlation potential fitting using an approach that relies on convex optimization---in particular, semidefinite programming (SDP). The SDP reformulation allows us to provably find the correlation potential, when the correlation potential is uniquely defined. It also allows us to use state-of-the-art numerical methods and software packages to compute the correlation potential in a robust fashion. Meanwhile, L-DMET improves the efficiency of the correlation potential fitting by replaces the global fitting procedure with several local correlation potential fitting procedures for each fragment. Moreover, we have shown that under certain natural conditions, the fixed points of L-DMET coincide with the original DMET. We demonstrate the accuracy, efficiency, and robustness of the L-DMET method by testing on Hubbard models and the hydrogen chain.

The question of whether the correlation potential is uniquely defined is central to both DMET and L-DMET. We show that in order to obtain a unique correlation potential, a necessary condition is that $N_e \geq L_A$, i.e., that the total electron number is larger than the fragment size. In practice we observe that the correlation potential is indeed often (but not always) unique when $N_e \geq L_A$.  We remark that the issue of finding a unique correlation potential is particularly relevant now due to the recent progress of \textit{ab initio} DMET calculations \cite{cui2019efficient}, where the fragment size can be large due to the use of a large basis set. Hence a rigorous understanding of sufficient conditions for the uniqueness of the correlation potential, as well as practical remedies when the correlation potential fails to be unique, are important issues that we shall consider in future work.

\section*{Acknowledgments:} 
This work was partially supported by the Air Force Office of Scientific Research under award number FA9550-18-1-0095 (X.W., M.L., Y.T., L.L.),  by the Department of Energy under Grant No. DE-SC0017867 (X.W. and L.L.),  by the Department of Energy CAMERA program (Y.T. and L.L.), by the National Science Foundation Graduate Research Fellowship Program under grant DGE-1106400 (M.L.), and by the National Science Foundation under Award No. 1903031 (M.L.).   We thank Berkeley Research Computing (BRC), Google Cloud Platform (GCP), and National Energy Research Scientific Computing Center (NERSC) for computing resources. We thank Garnet Chan for helpful discussions.

\bibliographystyle{achemso}
\bibliography{dmet_robust}

\appendix

%
%

\section{Uniqueness of the correlation potential}
\label{app:unique}

Here we demonstrate that the condition $N_e \geq L_A$ as in Assumption \ref{assump:nondegen} is necessary for the correlation potential to be unique. 
Suppose  $N_e < L_A$ and there exists  $u \in \mathcal{S}^0$ such that $f+u$ is gapped. Then let $u_1, \ldots, u_{N_e}$ be eigenvectors 
of $f+u$ spanning the occupied subspace, so that $D:= \mathcal{D}(f+u,N_e) = \sum_{i=1}^{N_e} u_i u_i^*$. 
Then let $v_1^x,\ldots,v_{N_e}^x \in \mathbb{C}^{L_A}$ be defined via $v_i^x = \left( \Phi_x^{\mathrm{frag}} \right)^\dagger u_i$ 
as the components of the $u_i$ within an arbitrary fragment $x$. Since $N_e < L_A$, there exists
some vector $w^x \in \mathbb{C}^{L_A}$ with $\Vert w^x \Vert = 1$ which is orthogonal to all of the $v_i^x$. Let 
$W =  \left(\Phi_x^{\mathrm{frag}}\right) (w^x) (w^x)^\dagger \left(\Phi_x^{\mathrm{frag}}\right)^\dagger$. Then by construction, all the $u_i$ are in the null space of $W$ for $i=1,\ldots, N_e$. Hence the $u_i$ are eigenvectors of $f + u + \tau(W-I)$ for $i=1,\ldots, N_e$, all $\tau \in \R$. Note that $\tau (W-I) \in \mathcal{S}^0$, and since $f+u$ is gapped, 
when $\tau$ is sufficiently small, we have $\mathcal{D}(f+u + \tau(W-I),N_e) = \sum_{i=1}^{N_e} u_i u_i^* = D$, contradicting 
uniqueness.

\section{Obtaining bath orbitals from the low-level density matrix }\label{sec:bath}

The global low-level density matrix, obtained from the decomposition in Eq. \eqref{eqn:c_ll} takes the form
\begin{equation}
D^{\text{ll}}=\left(\begin{array}{cc}{U_{A} \Sigma_{A}^{2} U_{A}^{\dagger}} & {U_{A} \Sigma_{A} \Sigma_{B} U_{B}^{\dagger}} \\ {U_{B} \Sigma_{B} \Sigma_{A} U_{A}^{\dagger}} & {U_{B} \Sigma_{B}^{2} U_{B}^{\dagger}+U_{\text{core}} U_{\text{core}}^{\dagger}}\end{array}\right):=
  \begin{pmatrix}
    D_{11} & D_{12}\\
    D_{21} & D_{22}
  \end{pmatrix}.
\label{eqn:dm_ll}
\end{equation}
where $D_{11}$ corresponds the fragment $x$ only. Then the CS decomposition \eqref{eqn:c_ll}, and hence the bath and core orbitals can also be identified from $D^{\text{ll}}$ directly. The eigenvalue decomposition of $D_{11}$ directly gives 
\begin{equation}
  \label{eq:8}
  D_{11} = U_A\Sigma_A^2U_A^\dagger .
\end{equation}
The bath-fragment density matrix can be written as
\begin{equation}
  \label{eq:15}
  D_{21}=C_BC_A^\dagger=U_B\Sigma_B\Sigma_AU_A^\dagger.
\end{equation}
The unitary matrix $U_B$ can be calculated by normalizing all the columns of the matrix $D_{21}U_A$, since $$U_B\Sigma_B\Sigma_A=D_{21}U_A.$$ The diagonal elements of $\Sigma_A\Sigma_B$ are the corresponding norms of the columns. As a result, $\Sigma_B$ is also obtained with the known $\Sigma_A$ in \eqref{eq:8}. Therefore we obtain the bath orbitals. Once the bath orbitals are obtained, the core orbitals can be obtained from the following relation
\begin{equation}
    U_{\text{core}} U_{\text{core}}^{\dag}= D_{22} - U_B\Sigma_B^2U_B^\dagger.
\end{equation}

\section{Proof of Proposition~\ref{prop:convex}}
\label{app:convex}

Heuristically, the idea for proving Proposition \ref{prop:convex} is that the first-order 
 optimality conditions for the optimization problem of Eq.~\eqref{eq:conjugate} (assuming differentiability 
 at the optimizer) are precisely $\nabla_{u_x} F(u) = D_{x}^{\mathrm{hl,frag}}$, i.e., equivalent 
 to exact fitting. However, some care is required when $F$ is singular at the optimizer.

We think of $F$ as a function on $(N_{\mathrm{f}})$-tuples of
$L_{A}\times L_{A}$ (Hermitian) matrices $(u_{x})_{x=1}^{N_{\mathrm{f}}}=(u_{1},\ldots,u_{N_{\mathrm{f}}})$.
This domain is identified with $\mathcal{S}$ as a slight abuse of notation.
As above we denote $u=\bigoplus_{x=1}^{N_{\mathrm{f}}}u_{x}$,
but by some abuse of notation we will also identify $u$ with $(u_{x})_{x=1}^{N_{\mathrm{f}}}=(u_{1},\ldots,u_{N_{\mathrm{f}}})$. 


Since we are given diagonal blocks $D_{x}^{\mathrm{hl},\mathrm{frag}}$
that we want to fit by choice of correlation potential blocks $u_{x}$,
we want to invert the gradient of $F$. We roughly understand that the gradient of $\nabla F=(\nabla_{u_{x}}F)_{x=1}^{N_{\mathrm{f}}}$
is invertible (up to shifting by a scalar matrix), with inverse specified
by the gradient of the concave conjugate or Legendre-Fenchel transform
$F^{*}$. But since $F$ is not differentiable everywhere, in fact
the \emph{supergradient}~\cite{rock} mapping $\partial F$ must be considered.
Under this mapping, each singular point $(u_{x})_{x=1}^{N_{\mathrm{f}}}$
of $F$ maps to all $(P_{x})_{x=1}^{N_{\mathrm{f}}}$ lying in
the supergradient set of $F$ at $(u_{x})_{x=1}^{N_{\mathrm{f}}}$,
i.e., all $(P_{x})_{x=1}^{N_{\mathrm{f}}}$ such that 
\[
F(v)\leq F(u)+\sum_{x}\Tr[P_{x}(v_{x}-u_{x})]
\]
 for all $v\in S$.

The set of optimizers of~\eqref{eq:conjugate}
 is precisely the set $\partial F^* (P)$~\cite{rock}. Moreover we have  $P \in \partial F(u)$ if and only if $u \in \partial F^* (P)$~\cite{rock}. Hence provided that $P$ is in the supergradient image of $F$, the set of optimizers of~\eqref{eq:conjugate} is 
 nonempty, and any element $u^\star$ satisfies $P \in \partial F(u^\star)$. Moreover, if $f + u^\star$ is gapped, then 
 as previously discussed $F$ is differentiable at $u^\star$, i.e., the subgradient is a singleton, and $\nabla F (u^\star) = P$, i.e., $u^\star$ attains exact fitting according to $P$. Finally, if $u^\star$ is the unique optimizer, then it follows that there does not exist $u \neq u^\star$ such that $P \in \partial F(u)$. Hence if $u^\star$ is the unique optimizer and $f+u^\star$ is gapless, then there is no correlation potential yielding an exact fit. 
 
%
%

 Then to complete the proof it suffices to show that our assumptions on $(D_x^{\mathrm{hl,frag}})_{x=1}^{N_{\mathrm{f}}}$ (i.e., that $0 \prec D_x^{\mathrm{hl,frag}} \prec I_{L_A}$ and  $\sum_x \Tr [D_x^{\mathrm{hl,frag}}] = N_e$) imply that $(D_x^{\mathrm{hl,frag}})_{x=1}^{N_{\mathrm{f}}}$ lies in the supergradient image of $F$. To
understand the supergradient image of $F$ and how to construct the
correlation potential $u$ more explicitly, we must study the concave
conjugate $F^{*}$.

Recall that the effective domain $\mathrm{dom}(F^{*})$ of $F^{*}$
is defined as the set of all points for which $F^{*}>-\infty$. The
relative interior (i.e., the interior of the effective domain within its affine hull~\cite{rock})
of the effective domain coincides with the supergradient
image of $F$~\cite{rock}, so we want to understand it.

 
To this end we shall concoct an alternate formula for $F^{*}$. First recall that $F^{**}=F$,
i.e., 
\[
F(u)=\inf_{(P_{x})_{x=1}^{N_{\mathrm{f}}}\in\mathrm{dom}(F^{*})}\left\{ \sum_{x=1}^{N_{\mathrm{f}}}\Tr[P_{x}u_{x}]-F^{*}(P_{1},\ldots,P_{N_{\mathrm{f}}})\right\} .
\]
 Meanwhile observe that for $A$ Hermitian, 
\[
\mc{E}_{N_{e}}(A)=\inf\left\{ \Tr[AP]\,:\,0\preceq P\preceq I_{L},\ \Tr[P]=N_{e}\right\} ,
\]
 so applying this result to $F(u)=\mc{E}_{N_{e}}(f+u)$, we see that
\begin{eqnarray*}
F(u) & = & \mc{E}_{N_{e}}(f+u)\\
 & = & \inf_{P^{\dagger}=P}\left\{ \sum_{x=1}^{N_{\mathrm{f}}}\Tr[P_{x}u_{x}]-G(P)\right\} \\
 & = & \inf_{(P_{x})\,\mathrm{Hermitian}}\left\{ \sum_{x=1}^{N_{\mathrm{f}}}\Tr[P_{x}u_{x}]-\sup_{P\,:\,P_{x}=[P]_{x}\,\forall x}G(P)\right\} 
\end{eqnarray*}
 where 
\[
G(P)=\begin{cases}
-\Tr[tP], & 0\preceq P\preceq I_{L},\ \Tr[P]=N_{e}\\
-\infty, & \mathrm{otherwise}.
\end{cases}
\]
 But consequently $F=g^{*}$, where 
\[
g(P_{1},\ldots,P_{N_{\mathrm{f}}}) =  \sup_{P\,:\,P_{x}=[P]_{x}\,\forall x}G(P)
 = -\inf_{0\preceq P\preceq I_{L}\,:\,\Tr[P]=N_{e},\,[P]_x = P_x \forall x}\Tr[tP].
\]

Then it its clear that 
\[
\mathrm{dom}(F^{*})=\left\{ (P_{1},\ldots,P_{N_{\mathrm{f}}})\,:\ 0\preceq P_{x}\preceq I_{L_{A}}\ \mathrm{for}\ x=1,\ldots,N_{\mathrm{f}},\ \ \sum_{x=1}^{N_{\mathrm{f}}}\Tr[P_{x}]=N_{e}\right\} .
\]
 Hence the relative interior of the effective domain is given by 
\[
\mathrm{relint}\,\mathrm{dom}(F^{*})=\left\{ (P_{1},\ldots,P_{N_{\mathrm{f}}})\,:\ 0\prec P_{x}\prec I_{L_{A}}\ \mathrm{for}\ x=1,\ldots,N_{\mathrm{f}},\ \ \sum_{x=1}^{N_{\mathrm{f}}}\Tr[P_{x}]=N_{e}\right\} .
\]
Our assumption on $(D_x^{\mathrm{hl,frag}})_{x=1}^{N_{\mathrm{f}}}$ was precisely that it lies in this set, so 
$(D_x^{\mathrm{hl,frag}})_{x=1}^{N_{\mathrm{f}}}$ lies in the supergradient image of $F$, and the 
proof is complete.

\section{Proof of Proposition~\ref{prop:sdp}}
\label{app:sdp}
Recall that for fixed $P=(P_{1},\ldots,P_{N_{\mathrm{f}}})$ satisfying
$0\prec P_{x}\preceq I_{L_{A}}$ for all $x$ and $\sum_{x}\Tr[P_{x}]=N_{e}$,
we want to solve 
\[
\inf_{u\in S^{0}}\left[\sum_{x}\Tr[P_{x}u_{x}]-F(u)\right].
\]
 Recall that 
\[
F(u)=F(u_{1},\ldots,u_{N_{\mathrm{f}}})=\mc{E}_{N_{e}}[h+u],
\]
 and $\mc{E}_{N_{e}}$ indicates sum of lowest $N_{e}$ eigenvalues.
We will write $F(u)$ as the optimal value of a suitable concave maximization
problem and plug this into the above convex minimization problem to
derive an SDP equivalent to what we want to solve.

First we observe that for any symmetric $A$ and any $m$, we can
write $\mc{E}_{m}(A)$ as the optimal value of the convex \emph{minimization
}problem: 
\[
\mc{E}_{m}(A)=\inf\left\{ \Tr(AX)\,:\,\Tr(X)=m,\,0\preceq X\preceq I\right\} .
\]
 Then we will derive the dual of this minimization problem to write
$\mc{E}_{m}(A)$ as the optimal value of a concave maximization problem.
To wit, write the Lagrangian, 
\begin{eqnarray*}
\mathcal{L}(X,Y,Z,\alpha) & = & \Tr(AX)-\Tr(YX)-\Tr(Z[I-X])-\alpha(\Tr(X)-m)\\
 & = & \Tr([A-Y+Z-\alpha I]X)+\alpha m-\Tr(Z)
\end{eqnarray*}
 where the domain is defined by $X$ symmetric, $Y\succeq0$, $Z\succeq0$,
$\alpha\in\R$. Then carry out the minimization over $X$ to derive
the dual problem 
\begin{eqnarray*}
 & \underset{Y\succeq0,Z\succeq0,\alpha\in\R}{\mathrm{maximize}} \quad\quad & \alpha m-\Tr(Z)\\
 & \mbox{subject to} \quad\quad & A-Y+Z-\alpha I\succeq0.
\end{eqnarray*}
 Evidently it is optimal to choose $Y=0$, hence we have the equivalent
program 
\begin{eqnarray*}
 & \underset{Z\succeq0,\alpha\in\R}{\mathrm{maximize}} \quad\quad & \alpha m-\Tr(Z)\\
 & \mbox{subject to}\quad\quad & A+Z-\alpha I\succeq0.
\end{eqnarray*}
The optimal value is equal to $\mc{E}_{m}(A)$ by strong duality,
i.e., we can write
\[
\mc{E}_{m}(A)=\max\left\{ \alpha m-\Tr(Z)\,:\,A+Z-\alpha I\succeq0,\ Z\succeq0,\ \alpha\in\R\right\} .
\]

Applying this result for $A=h+u$, we see that we can rephrase our
original optimization problem as
\begin{eqnarray*}
 & \underset{u\in S^{0},\,Z\in\mathbb{C}^{M\times M}\,\mathrm{Hermitian},\,\alpha\in\R}{\mathrm{minimize}} \quad\quad & \sum_{x=1}^{N_{\mathrm{f}}}\Tr[P_{x}u_{x}]-\alpha N_{e}+\Tr(Z)\\
 & \mbox{subject to} \quad\quad & h+u+Z-\alpha I\succeq0\\
 &  & Z\succeq0.,
\end{eqnarray*}
 as was to be shown.

\section{Proof of Proposition \ref{prop:equiv}}\label{app:equivalency}
We first consider a fixed point of DMET denoted by $u^{\star}$, which solves Eq. \eqref{eqn:dmet_selfconsistency} with $\mathfrak{F}=\mathfrak{F}^{\text{DMET}}$. Then for any $x$
\[
D_x^\text{hl,frag}=E^{\top}\Phi_x^{\dag}\mathcal D(f+u^{\star}, N_e)\Phi_x E,
\]
where as before $E=(I_{L_A},0_{L_A\times L_A})^\top$ If we can further show that  for any $x$,
\begin{equation}
E^{\top}\Phi_x^{\dag}\mathcal D(f+u^{\star}, N_e)\Phi_x E=E^\top \mathcal D(\Phi_x^{\dag}(f+u^{\star})\Phi_x, L_A)E.
\label{eqn:dm_ll_equiv}
\end{equation}
then by the uniqueness of the local correlation fitting we have $\wt{u}_x=0$. Therefore $u^{\star}$ is a fixed point problem of the L-DMET.

Without loss of generality, we assume fragment $x$ consists of orbitals
$\{1,2,\ldots,L_{A}\}$. Using the notation in Eq. \eqref{eqn:c_ll}, 
the basis transformation matrix is
\[
U=\left(\begin{array}{cccc}
I & 0 & 0 & 0\\
0 & U_{B} & U_{\text{core}} & U_{\text{vir}}
\end{array}\right)\in\CC^{L\times L}.
\]

It can be obtained via
\begin{equation}
\underset{C\in \CC^{L\times N_e},C^{\dagger}C=I_{N_{e}}}{\mathrm{minimize}\quad}\mathrm{Tr}[C^{\dagger}(f+u^{\star})C],
\label{eq:low_level_opt}
\end{equation}
and with respect to the new basis defined by $U$, Eq. \eqref{eq:low_level_opt} becomes
\begin{equation}
\underset{\tilde{X}\in \CC^{L\times N_e},\tilde{X}^{\dagger}\tilde{X}=I_{N_{e}}}{\mathrm{minimize}\quad}\mathrm{Tr}[\tilde{X}^{\dagger}U^{\dagger}(f+u^{\star})U\tilde{X}].
\label{eq:low_level_opt_proj}
\end{equation}
Using the decomposition \eqref{eqn:c_ll}, we have 
\begin{equation}
\tilde{X}=U^{\dagger}C=\left(\begin{array}{c}
U_{A}\Sigma_{A}V^{\dagger}\\
\Sigma_{B}V^{\dagger}\\
V_{\perp}^{\dagger}\\
0
\end{array}\right).\label{eq:low_level_optimizer}
\end{equation}
Now we constrain $\tilde{X}$ to take a more general form 
\[
\tilde{X}=\left(\begin{array}{c}
XV^{\dagger}\\
V_{\perp}^{\dagger}\\
0
\end{array}\right)\in \CC^{L\times N_e},
\]
where $X\in\mathbb{C}^{2L_{A}\times L_{A}}$ and $X^{\dagger}X=I_{L_{A}}$. Then we have
\[
\Tr[\tilde{X}^{\dagger}U^{\dagger}(f+u^{\star})U\tilde{X}]=\Tr[X^{\dagger}\Phi_{x}^{\dagger}(f+u^{\star})\Phi_{x}X]+\Tr[V_{\perp}^{\dagger}\Xi V_{\perp}],
\]
where $\Xi$ is the diagonal matrix consisting of the eigenvalues of core orbitals. Since the second term
on the right hand side does not depend on $X$, then $X=\left(\begin{array}{c}
U_{A}\Sigma_{A}\\
\Sigma_{B}
\end{array}\right)$ solves the following minimization problem
\[
\underset{X\in\CC^{2L_A\times L_A},X^{\dagger}X=I_{L_{A}}}{\mathrm{minimize}}\mathrm{Tr}X^{\dagger}\Phi_{x}^{\dagger}(f+u^{\star})\Phi_{x}X
\]
Therefore 
\[
E^\top \mathcal D(\Phi_x^{\dag}(f+u^{\star})\Phi_x, L_A)E=E^{\top}XX^{\dag}E=U_{A}\Sigma_{A}^{2}U_{A}^{\dagger}=E^{\top}\Phi_x^{\dag}\mathcal D(f+u, N_e)\Phi_x E .
\]
The last equality follows from Eq. \eqref{eqn:dm_ll}. 

Similarly if $u^{\star}$ is a fixed point of L-DMET, by Eq. \eqref{eqn:dm_ll_equiv} it is also a fixed point of DMET.

\section{Comparison of semidefinite programming and least squares fitting in 1D Hubbard model}\label{sec:comp_hubbard_1d}

\begin{figure}[!htp]
  \centering
  \includegraphics[scale=0.7]{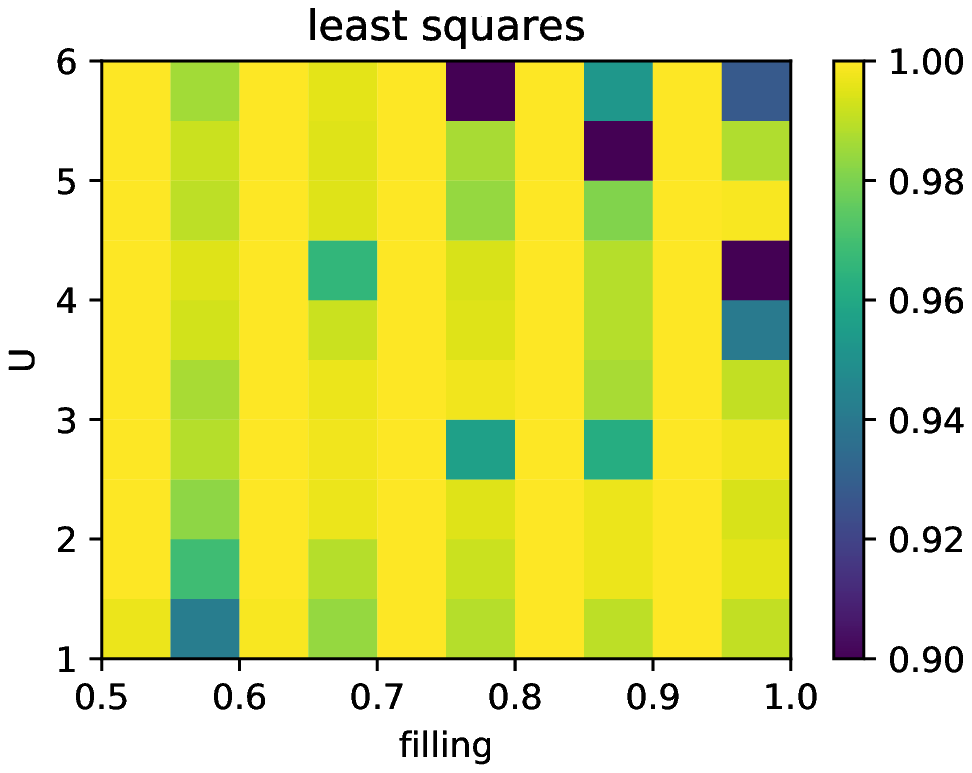}
  \includegraphics[scale=0.7]{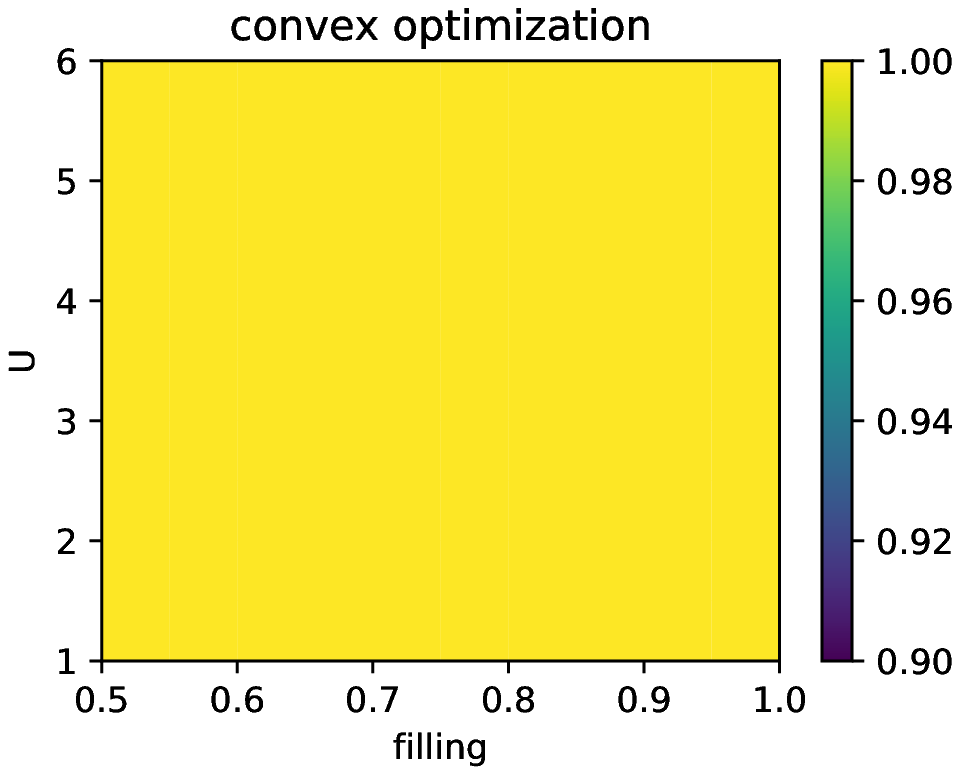}
  \caption{Success rates of the least squares (left) and convex optimization (right) approaches for the 1D Hubbard model.}
  \label{fig:success1d}
\end{figure}

To further evaluate the comparison between the SDP and least squares fitting, we repeat the analysis of their success rates following exactly the same procedure as outlined in section~\ref{sec:SDPcompare}, except that we now instead consider a 1D Hubbard model. In particular, we consider a 1D Hubbard chain of 40 sites with anti-periodic boundary condition, and we take fragments consisting of 2 sites. 
As shown in Fig. \ref{fig:success1d}, the least squares approach clearly performs better than it does on the 2D Hubbard Model. Nonetheless, the least squares frequently fails when the number of electrons is $24, 28, 32, 36$ and $40$. Meanwhile, the SDP approach enjoys a $100\%$ success rate on our test cases. The experiment for the 1D Hubbard model indicates again the SDP approach is more robust than the least squares approach. 
\end{document}